\documentclass[a4paper,reqno]{amsart}
\usepackage{geometry}
\usepackage{fullpage}
\usepackage[latin1]{inputenc}
\usepackage[T1]{fontenc}
\usepackage{amsfonts}
\usepackage{color}
\usepackage{epsfig}
\usepackage{epstopdf}
\usepackage{subfigure}
\usepackage{bm}
\usepackage{bbold}
\usepackage{amssymb}
\usepackage{amsmath}
\usepackage{amsthm}
\usepackage{graphicx}
\usepackage{subfigure}
\usepackage[dvipsnames]{xcolor}
\usepackage{afterpage}
\usepackage{booktabs}
\usepackage{siunitx}

\usepackage[colorlinks=true, linkcolor=blue, citecolor=cyan, urlcolor=green]{hyperref}\usepackage{bbm}
\usepackage{latexsym}
\usepackage{amsaddr}
\usepackage{booktabs}
\usepackage{fix-cm} 

\numberwithin{equation}{section}

\definecolor{light}{gray}{.9}

\def\be{\begin{equation}}
\def\ee{\end{equation}}
\def\bea{\begin{eqnarray}}
\def\eea{\end{eqnarray}}

\def\E{\mathbb{E}}
\def\ThetaB{\bm{\Theta}}

\newcommand{\nocontentsline}[3]{}
\newcommand{\tocless}[2]{\bgroup\let\addcontentsline=\nocontentsline#1{#2}\egroup}

\DeclareMathSymbol{\leqslant}{\mathalpha}{AMSa}{"36} 
\DeclareMathSymbol{\geqslant}{\mathalpha}{AMSa}{"3E} 
\DeclareMathSymbol{\eset}{\mathalpha}{AMSb}{"3F}     

\def\P{\mathbb{P}}
\def\E{\mathbb{E}}
\def\A{\bm{A}}

\title{A dynamic network model with persistent links and node-specific latent variables, with an application to the interbank market}
\date{\today}

\author{P. Mazzarisi}
\email{piero.mazzarisi@sns.it}
\address{Scuola Normale Superiore, Pisa, Italy} 
\author{P. Barucca}
\email{paolo.barucca@bf.uzh.ch}
\address{Department of Banking and Finance, University of Zurich, Switzerland\\
London Institute for Mathematical Sciences, London, UK}
\author{F. Lillo}
\email{fabrizio.lillo@unibo.it}
\address{Department of Mathematics, University of Bologna, Italy}
\author{D. Tantari}
\email{daniele.tantari@sns.it}
\address{Scuola Normale Superiore, Pisa, Italy} 

\keywords{Dynamic networks, Markov dynamics, autoregressive models, fitness model, link persistence, expectation-maximization, preferential trading, link prediction}

\begin{document}
\maketitle

\begin{abstract}
We propose a dynamic network model where two mechanisms control the probability of a link between two nodes: (i) the existence or absence of this link in the past, and (ii) node-specific latent variables (dynamic fitnesses) describing the propensity of each node to create links. Assuming a Markov dynamics for both mechanisms, we propose an Expectation-Maximization algorithm for model estimation and inference of the latent variables. The estimated parameters and fitnesses can be used to forecast the presence of a link in the future. We apply our methodology to the e-MID interbank network for which the two linkage mechanisms are associated with two different trading behaviors in the process of network formation, namely preferential trading and trading driven by node-specific characteristics. The empirical results allow to recognise preferential lending in the interbank market and indicate how a method that does not account for time-varying network topologies tends to overestimate preferential linkage.
\end{abstract}

\section*{Introduction}

In recent years there has been a growing interest in the study of complex networks \cite{newman2001,holme2012}. One of the reasons is that many natural and artificial systems are characterized by the presence of a sparse structure of interactions, i.e. only a small fraction of the possible pairs of elements mutually interact (at least at each time). Thus the topology of the network of interactions plays an important role in understanding the aggregate behavior of many complex systems. Moreover most of the investigated systems evolve over time and the structure of the network is generically not constant but new links are formed and old ones are destroyed at each time. Understanding and modeling network dynamics is therefore of paramount importance in many disciplines as testified by the recent literature (see below for a review).

From a modeling perspective there are several mechanisms that can lead to link formation and destruction.  The presence of a link may depend on present node properties of the system but also on previous network states. Consider as an example a trading network such as the interbank network studied in the empirical application below. The probability of a link between two nodes representing a transaction between the corresponding entities depends generically on the current supply and demand of the two entities, as well as on the existence (or absence thereof) of a link in the past between the two entities. The former driver is associated with node-specific properties (supply and demand) which evolve in time with their own dynamics, possibly dependent also on the network state. The latter driver, instead, is associated with a link specific property, namely the persistence of links which describes the tendency to interact with whom we have interacted with in the past.

The objective of this paper is to introduce a dynamic network model where both mechanisms are present and to propose a statistical estimation technique which allows to disentangle the importance of the two mechanisms for each link in the network. The estimation method is based on an Expectation-Maximization scheme for maximum likelihood estimation. As we will show, the estimation of the model allows also to forecast the probability of the existence of a specific link in the future when the model parameters and the past network history are known.

More specifically, in our model we define a Markovian process on link dynamics combined with an autoregressive model for the latent variables governing the link probability. There is a latent variable in each node and it is termed the fitness of the node. Thus at each time-step a link can be created - or not - either as a consequence of a copying process of the past link state or as a consequence of a random sampling whose probability depends on the current value of the latent variables of the two considered nodes. Clearly both mechanisms give rise to time correlation of the link state, even if its origin is quite different in the two cases. Being able to disentangle the role in link persistence due to explicit copy from the past or to fitness dynamics allows to identify genuine patterns of preferential linkage.  

As a specific application in this paper we study the interbank market, which is an important infrastructure of the financial system. Banks borrow and lend money in the interbank market to meet liquidity shortages or to allocate liquidity surpluses on a daily basis. The decision of whom to trade with is complex but certainly two aspects play an important role: first, the internal state (e.g. balance sheet, liquidity available or needed) of the bank and second the knowledge of the counterpart. Concerning this last aspect, all else being equal a bank will prefer to trade with someone who was a counterpart in the past, since lending money requires some trust on the borrower's solvability. This behavior is known as preferential trading \cite{weisbuch2000} and has been documented in many empirical papers \cite{cocco2009}. Our model is able to assess the importance of preferential trading between two banks when the (possibly time-varying) internal states of the two banks are taken into account. It is important to stress that the same argument can be made for social networks where  the copying mechanism favors extant links due to a minor social cost of entertaining new relationships.

\paragraph{{\bfseries Related Literature}}

The literature on statistical models of temporal networks is rapidly growing and there are several dynamic network models that have been investigated. Each model tries to capture different aspects of spatial and time dependencies in temporal networks and the main aims are the description of how network topology evolves through time and the prediction of links. From a mathematical point of view, there exist two principal approaches: (i) description of the graph dynamics using (generalized) Markov chains on network observables (ii) description of the graph dynamics using models of latent variables whose dynamics determines the evolution of the network topology. We briefly describe both approaches in the following.

Concerning the first stream of literature, a milestone study is represented by the work of Hanneke et al. \cite{hanneke2010} where Exponential Random Graphs (ERG) model has been adapted to temporal networks. The method is called Temporal Exponential Random Graphs (TERG) and aims to model several metrics involving two consecutive network snapshots in a similar fashion to ERG. Krivitsky and Handcock \cite{kriv2014} investigated further TERG by studying a specific parametrization allowing maximum likelihood estimation of the model. More recently, Peixoto and Rosvall \cite{peixoto2015} have proposed an extension of the Stochastic Block Model (SBM) to temporal networks by modeling a $n$-th order Markov chain with suitable transition probabilities which generate the time sequence of links. The aim is to select the most appropriate Markov order and number of communities. Zhang et al. \cite{zhangmoore2016} have proposed generalizations of a number of standard network models, including the classic random graph, the configuration model as well as SBM, to the case of dynamic networks in a similar fashion to \cite{peixoto2015}. Furthermore, in this stream of literature several works have focused on the problem of link prediction in time-evolving graphs. One example is the work of Richard et al. \cite{gaiffas2014} where authors propose to describe the time series of graph snapshots with a vector autoregressive process. Furthermore, they propose an efficient model estimation based on proximal methods. 

However, this first kind of approach has been criticized because it can not fully capture the time-varying patterns of the network structure. This opens to a second stream of literature which aims to describe these patterns with models having time-varying (latent) parameters that capture how network topology changes in time, see \cite{reviewlatent} for a review. A milestone work is represented by the study of Sarkar and Moore \cite{sarkar2005} that generalized the latent space model introduced in \cite{hoff2002} to dynamic networks. The dynamics of the network structure is modelled through random effects in a latent space. Sewell and Chen \cite{sewell2015} proposed a Markov Chain Monte Carlo (MCMC) algorithm to estimate the model parameters and latent positions of the nodes in the network. Durante and Dunson \cite{durante2014} proposed a further extension of the model in \cite{sarkar2005} by describing a latent space model for dynamic networks in which latent node positions evolves in time via stochastic differential equations. They introduced also an efficient MCMC algorithm for Bayesian inference to learn model parameters \cite{durante2016}. Giraitis et al. \cite{giraitis2016} proposed a novel methodology for dynamic modeling of temporal networks with application to interbank networks. They describe the link dynamics with a Tobit model, allowing for deterministic or stochastic time-varying parameters that take into account the possibility of structural changes in network dynamics. Brauning and Koopman \cite{koopman2016} have applied the dynamic factors model to the case of dynamic networks. Depending on the number of factors, the model allows to reduce the dimensionality of the problem and to describe cross-sectional dependencies in network data. Lee at al. \cite{lee2017} have recently introduced a generalization of TERG, called varying-coefficient exponential random graph model, that characterizes the evolution of network topologies through smoothly time varying parameters whose dynamics can capture temporal heterogeneity of dynamic networks. Finally, in this stream of literature we can also include all the generalizations \cite{yang2011detecting,xu2014dynamic,xu2015stochastic,ghasemian2016,blmt2017} of SBM  which account for time evolving community memberships and/or link persistence.  However, these works are more focused on the  problem of community detection when dynamic effects are considered.

The methodology we propose in this paper exploits both of the aspects that characterize the two streams of literature. From one side, we describe link persistence coming from the mechanism of copying from the past by modeling a Markov chain for link stability, i.e. the tendency of a link that does (or does not) exist at time $t-1$ to continue existing (or not existing) at time $t$, similarly to \cite{hanneke2010}. From the other side, we describe the stochastic dynamics of node-specific latent variables that we call fitnesses, with a similar aim of \cite{koopman2016,lee2017}. The node fitness describes the tendency of a node in creating links and its evolution determines how the degree of the node changes in time. 

From the point of view of generative network models, link stability tends to capture preferential linkage mechanism between the nodes of the network while the fitness dynamics accounts for the evolving network topology. The main goal of this work is to disentangle the two temporal patterns generated by the two linkage mechanisms in network data. Hence, we apply our methodology to the financial network of electronic Market of Interbank Deposit (e-MID) where the two linkage mechanisms are associated with two different trading behaviors, i.e. random and preferential trading \cite{tumminiello2015}, in the process of network formation.

The remainder of this paper is organized as follows. In Section \ref{methods} we describe three different models of temporal networks. In Section \ref{EM} we present a novel Expectation-Maximization algorithm for model estimation and in Section \ref{montecarlo} we run a Monte Carlo exercise to assess the goodness of fit of our estimation method. In Section \ref{empirical} we apply our methodology to the network of the electronic Market of Interbank Deposit. Finally, we conclude with a discussion of our method and open areas for future research in Section \ref{conclusion}.

\section{The models}\label{methods}

In this Section, we introduce three models of temporal networks: (i) in the first one, the presence or absence of a link can be a copy of the past with a given probability or can be sampled according to a Bernoulli marginal distribution; (ii) in the second model, each graph snapshot does not have an explicit dependence from the past snapshots but the link probability depends on node-specific latent dynamical variables, i.e. the node fitnesses, which evolve stochastically in time with memory of past information; (iii) the third network model combines the copying mechanism of the first model with dynamic node fitnesses of the second model.

We define a temporal network as a time series of graphs, that is the set $(V,\{\A^t\}^{t=0,1,...,T})$ with $|V|=N$ nodes and adjacency matrices $\{\A^t\}^{t=0,1,...,T}$. A network snapshot is the observed graph at a given time $t$ and is described by the adjacency matrix $\A^t$ which has entry $A_{ij}^t=1$ if the edge from node $i$ to node $j$ is present at time $t$ and zero otherwise. In our models, we exclude graphs with self loops, i.e. the diagonal of $\A^t$ is null for all $t$. The adjacency matrix can be symmetric (undirected graphs) or not (directed ones). In the following, we refer to the undirected case for notational simplicity. The generalization is straightforward and is indeed used in the empirical analysis of the interbank market of Section \ref{empirical}.

In our framework, a temporal network is the observable of the following hidden Markov chain:
\begin{equation}\label{hmc}
\begin{cases}
\P(\Theta^{t}|\Theta^{t-1},\bm{\Phi})&=h(\Theta^t,\Theta^{t-1},\bm{\Phi})\\
\P(\A^t|\A^{t-1},\Theta^t,\bm{\beta})&=g(\A^t,\A^{t-1},\Theta^t,\bm{\beta})
\end{cases}
\end{equation}
where $\{\Theta^t\}^{t=0,1,...,T}$ represents the set of dynamic parameters, which are also called latent variables of the Markov chain. Their dynamics is determined by the one-step transition probability  $h$, whereas $g$ represents the likelihood for the network snapshot at time $t$ given the information about the previous network snapshot, as well as the latent variables $\Theta^t$. Finally $\Pi\equiv\{\bm{\beta},\bm{\Phi}\}$ represents the set of static parameters.

Since the Markov chain in Eq. \ref{hmc} has a high-dimensional set of parameters, we reduce the dimensionality by assuming that the node-specific latent variables evolve independently and that there are no explicit spatial correlations among links. Nevertheless, spatial correlations between links are implicitly induced by the latent dynamics.

As mentioned above, in the following, we consider three different specifications of Eq. \ref{hmc}.

\subsection{Discrete AutoRegressive Graphs (DAR$(1)$)}
We model link stability with the following discrete autoregressive process,
\begin{equation}\label{dar1}
A_{ij}^t=V_{ij}^tA_{ij}^{t-1}+(1-V_{ij}^t)Y_{ij}^t\:\:\:\forall i,j=1,...,N \:\:\mbox{and}\:\: j>i
\end{equation}
where $V_{ij}^t\sim\mathcal{B}(\alpha_{ij})$ with $\alpha_{ij}\in[0,1]$, $Y_{ij}^t\sim\mathcal{B}(\chi_{ij})$ with $\chi_{ij}\in[0,1]$ and $\mathcal{B}$ indicates the Bernoulli distribution. In the process of Eq. \ref{dar1}, the value of $A_{ij}^t$ is copied from the past value with probability $\alpha_{ij}$ or obtained by tossing a coin according to the marginal distribution $\mathcal{B}(\chi_{ij})$ with probability $1-\alpha_{ij}$. Highly persistent links (or no-links) are described by high values of $\alpha_{ij}$. As a consequence, networks characterized by high values of $\bm{\alpha}\equiv\{\alpha_{ij}\}_{i,j=1,...,N}$ tend to preserve the past structure through time.

The Markov chain described by Eq. \ref{dar1} is the first order process DAR(1), belonging to the more general class of discrete autoregressive processes DAR(p) \cite{jacobs1978}. Here, we do not consider the hidden dynamics associated with the latent variables.

Hence, the specification of Eq. \ref{hmc} for this model is the following,
\begin{equation}\label{Amc}
\P(\A^t|\A^{t-1},\bm{\alpha},\bm{\chi})=\prod_{i,j>i}\P(A_{ij}^t|A_{ij}^{t-1},\alpha_{ij},\chi_{ij})=\prod_{i,j>i}\left(\alpha_{ij}\mathbb{I}_{A_{ij}^{t}A_{ij}^{t-1}}+(1-\alpha_{ij})\chi_{ij}^{A_{ij}^{t}}(1-\chi_{ij})^{1-A_{ij}^{t}}\right),
\end{equation}
where $\mathbb{I}_{A_{ij}^{t}A_{ij}^{t-1}}$ is the indicator function taking value equal to $1$ if $A_{ij}^t=A_{ij}^{t-1}$ and zero otherwise. Eq. \ref{Amc} describes $\binom{N}{2}$ independent Markov chains for each link. This model of temporal networks is fully determined by the $N(N-1)$ parameters $\{\bm{\alpha},\bm{\chi}\}\equiv\{\alpha_{ij},\chi_{ij}\}_{i=1,...,N;j>i}$ and we estimate them by maximum likelihood method. 

The persistence pattern of this model can be quantified by the autocorrelation functions (ACF) of the links. It is the one of a standard autoregressive process AR(1) but with non negative autoregressive coefficient $\alpha_{ij}$, i.e. the DAR(1) graph model is able to describe only non negative ACF. The generalization of this model to directed networks is simply obtained by considering not symmetric adjacency matrices.

\subsection{Temporally Generalized Random Graphs (TGRG)} The second model is a generalization of the fitness network model \cite{caldarelli2002,garlaschellifit2004} to a dynamic setting that accounts for time evolving node fitness. Fitness is a node property determining its capability of creating links. We assume that each node $i$ is characterized by the fitness $\theta_i$ which evolves in time by following a covariance stationary autoregressive process AR(1),
\begin{equation}\label{ar1}
\theta_i^t=\phi_{0,i}+\phi_{1,i}\theta_i^{t-1}+\epsilon_{i}^t\:,\:\:\:\:\forall i=1,...,N
\end{equation}
where $\phi_{0,i}\in\mathbb{R}$, $|\phi_{1,i}|<1$ and the i.i.d. variables $\epsilon_i^t\sim\mathcal{N}(0,\sigma_i^2)$. This choice is consistent with the Markovian assumption in Eq. \ref{hmc}. Moreover, the hidden node state $\theta_{i}^t$ evolves in $\mathbb{R}$ between timesteps, but large changes are unlikely because of the Gaussian transition probabilities. This is consistent with the idea that the network topology changes smoothly in time. Finally, assuming a Gaussian transition probability represents a simplification for model estimation.

The conditional probability for the network at time $t$ is
\begin{equation}\label{pAfitness}
\P(\A^t|\Theta^t)=\prod_{i,j>i}\frac{e^{A_{ij}(\theta_i^t+\theta_j^t)}}{1+e^{(\theta_i^t+\theta_j^t)}},
\end{equation}
where $\Theta^t\equiv\{\theta^t_i\}_{i=1,...,N}$ is the vector of time-varying parameters. In Eq. \ref{pAfitness} we assume that each link is independently sampled and the probability of a link between node $i$ and node $j$ at time $t$ is determined by the corresponding $\theta_i^t$ and $\theta_j^t$. The larger is $\theta_i^t$, the larger is the probability for all links incident to node $i$.

We refer to this model as Temporally Generalized Random Graphs (TGRG) and the specification of Eq. \ref{hmc} for the TGRG is the following,
\begin{equation}\label{tgrg}
\begin{cases}
\P(\theta_i^t|\theta_{i}^{t-1},\bm{\Phi}_i)&=f(\theta_i^t|\phi_{0,i}+\phi_{1,i}\theta_i^{t-1},\sigma_i^2)\:\:\:\forall i=1,...,N\\
\P(\A^t|\Theta^t)&=\prod_{i,j>i}\P(A_{ij}^t|\theta_{i}^t,\theta_{j}^t)=\prod_{i,j>i}\frac{e^{A^t_{ij}(\theta_i^t+\theta_j^t)}}{1+e^{(\theta_i^t+\theta_j^t)}}
\end{cases}
\end{equation}
and $\P(\Theta^{t}|\Theta^{t-1},\bm{\Phi})=\prod_{i=1}^N\P(\theta_i^t|\theta_{i}^{t-1},\bm{\Phi}_i)$ according to the hypothesis of independence, where $f(\theta_i^t|\phi_{0,i}+\phi_{1,i}\theta_i^{t-1},\sigma_i^2)$ is the density of normal variable with mean $\phi_{0,i}+\phi_{1,i}\theta_i^{t-1}$ and variance $\sigma_i^2$. The set of static parameters is 
$\bm{\Phi}\equiv\{\bm{\Phi}_i\}_{i=1,...,N}$ with $\bm{\Phi}_i\equiv\{\phi_{0,i},\phi_{1,i},\sigma_i\}$.

The TGRG model is fully determined by the $3\times N$ static parameters $\bm{\Phi}$. In the next Section we propose an Expectation-Maximization scheme to estimate the model parameters and time-varying parameters. It alternates between an Expectation step where we fit the time-varying parameters $\{\Theta^t\}^{t=0,1,...,T}$ and the Maximization step where we maximize the log-likelihood of the static parameters conditional on the expectations $\{\hat{\Theta}^t\}^{t=0,1,...,T}$. 

Time autocorrelated node fitnesses may induce link persistence. In fact, the probability of a link between two specific nodes $e^{(\theta_i^t+\theta_j^t)}/ (1+e^{(\theta_i^t+\theta_j^t)})$ is persistent if $\theta_i^t$ and $\theta_j^t$ are autocorrelated. Note that link persistence occurs as a consequence of node properties. For TGRG, the two-point distribution function for lagged links and the ACF of link state can be semi-analytically computed (see the Appendix).

The generalization of the TGRG model for directed networks can be accomplished by distinguishing between the out-degree and the in-degree and by introducing two fitnesses for each node $i$, i.e. $\theta_i^{t,out}$ and $\theta_i^{t,in}$. The probability of a link from node $i$ to node $j$ at time $t$ given the latent variables $\theta_i^{t,out}$ and $\theta_j^{t,in}$ is $\P(A_{ij}^t|\theta_i^{t,out},\theta_j^{t,in})=\frac{e^{A_{ij}^t(\theta_i^{t,out}+\theta_j^{t,in})}}{1+e^{(\theta_i^{t,out}+\theta_j^{t,in})}}$. Then, everything follows similarly to the undirected case with the exception that $\P(\A^t|\Theta^t)$ is invariant under a linear transformation for the hidden node states: $\theta_i^{t,out}\mapsto\theta_i^{t,out}+c_t\:\:\:\forall i=1,...,N$, $\theta_j^{t,in}\mapsto\theta_j^{t,in}-c_t\:\:\:\forall j=1,...,N$, where $\{c_t\}_{t=0,1,...,T}\in\mathbb{R}^{T+1}$. This symmetry arises because the total number of outgoing links has to be equal to the total number of ingoing links at each time. This degeneracy can be simply removed by taking one of the fitnesses as constant in time.

Finally, let us notice that we can interpret TGRG as an extension of Exponential Random Graphs (ERG) \cite{parknewman2004} to the dynamic case. ERG ensembles are probability distributions of networks obtained by maximizing the Shannon entropy under some constraints on the average value of a set of network observables. If this set is the degree sequence, the Lagrange multipliers of the entropy constrained optimization can be directly linked to the latent variables of our model. Differently from other dynamic extension of ERG (see for example \cite{hanneke2010}) where dynamical (i.e. two-time) observables are used as constraints, here we choose a dynamical model for the latent variable, namely the AR(1) process, and introduce an estimation method for them.

\subsection{Discrete AutoRegressive Temporally Generalized Random Graphs (DAR-TGRG)} The persistence pattern associated with the copying mechanism described by Eq. \ref{dar1} can coexist with the node fitnesses evolving in time according to Eq. \ref{ar1}. This can be captured by the following specification of the model in Eq. \ref{hmc},
\begin{equation}\label{dartgrg}
\begin{cases}
\P(\theta_i^t|\theta_{i}^{t-1},\bm{\Phi}_i)&=f(\theta_i^t|\phi_{0,i}+\phi_{1,i}\theta_i^{t-1},\sigma_i^2)\:\:\:\forall i=1,...,N\\
\P(\A^t|\A^{t-1},\Theta^t,\bm{\alpha})&=\prod_{i,j>i}\left(\alpha_{ij}\mathbb{I}_{A_{ij}^tA_{ij}^{t-1}}+(1-\alpha_{ij})\frac{e^{A_{ij}^t(\theta_i^t+\theta_j^t)}}{1+e^{(\theta_i^t+\theta_j^t)}}\right)
\end{cases}
\end{equation}
and $\P(\Theta^{t}|\Theta^{t-1},\bm{\Phi})=\prod_{i=1}^N\P(\theta_i^t|\theta_{i}^{t-1},\bm{\Phi}_i)$ according to the hypothesis of independence, where $f(\theta_i^t|\phi_{0,i}+\phi_{1,i}\theta_i^{t-1},\sigma_i^2)$ is the density of a normal variable with mean $\phi_{0,i}+\phi_{1,i}\theta_i^{t-1}$ and variance $\sigma_i^2$, $\alpha_{ij}\in[0,1]$ and $\bm{\alpha}\equiv\{\alpha_{ij}\}$ $\forall i,j=1,...,N$ with $\alpha_{ij}=\alpha_{ji}$ for undirected networks, $\bm{\Phi}\equiv\{\phi_{0,i},\phi_{1,i},\sigma_i\}_{i=1,...,N}$ with $\phi_{0,i}\in\mathbb{R}$, $|\phi_{1,i}|<1$ and $\sigma_i>0$ $\forall i$, and $\theta_i^t\in\mathbb{R}$ $\forall i,t$.

This model can be interpreted as a mixture of the two mechanisms, i.e. the one of copying the presence or absence of a link from the past with probability $\alpha_{ij}$ and the one of time evolving marginals described by the TRGR model with probability $1-\alpha_{ij}$. Let us stress that the temporal pattern generated by the fitness dynamics does not concern a specific link but it is a node property. Thus under this mechanism, links incident on the same node tend to have similar persistence properties. On the contrary, the persistence of the copying mechanism is a link property, and links incident on the same node can have very different persistence properties. The parameter $\alpha_{ij}$ disentangles the importance of these two effects in determining the dynamics of the link $(i,j)$.

The model in Eq. \ref{dartgrg} is fully determined by the $\binom{N}{2}$ parameters $\bm{\alpha}$ and the $3\times N$ parameters $\bm{\Phi}$, which can be estimated by the Expectation-Maximization algorithm we propose in the next Section. 

\section{Estimation method}\label{EM}
We now describe the procedure for the estimation of the DAR-TGRG model. We propose an Expectation-Maximization method based on a Bayesian inference approach. The estimation method for TGRG model is simply obtained by setting parameters $\alpha_{ij}\:\:\forall i,j=1,...,N$ equal to zero in the following equations. 

Let denote $\ThetaB\equiv\{\Theta^t\}^{t=1,...,T}$, $\A\equiv\{\A^t\}^{t=0,1,...,T}$ and $\Pi\equiv\{\bm{\Phi},\bm{\alpha}\}$. The Bayesian approach considers the posterior distribution of the latent variables
\begin{equation}\label{postHidden}
\P(\ThetaB|\A,\Pi)=\frac{\P(\A,\ThetaB|\Pi)}{\int [d\Gamma]\P(\A,\bm{\Gamma}|\Pi)}=Z_{\Pi}^{-1}\P(\A,\ThetaB|\Pi),
\end{equation}
where $[d\Gamma]$ represents the measure over the probability space for $\ThetaB$, for inferring a set of statistically significant fitnesses $\hat{\ThetaB}$ and the posterior distribution over the static parameters
\begin{equation}\label{postParameters}
\P(\Pi|\A)=\frac{\P(\Pi)}{\P(\A)}\int[d\Gamma]\P(\A,\Gamma|\Pi)\propto \P(\Pi)Z_{\Pi}
\end{equation}
to learn the most likely set of parameters $\hat{\Pi}$ given the data. Using smooth priors $\P(\Pi)$, $\hat{\Pi}$ is obtained by extremizing over $\Pi$ the log-likelihood $l(\Pi)\equiv\log \P(\Pi|\A)$, i.e. by solving the equations
\begin{equation}\label{Eqlearn}
\partial_\Pi l(\Pi)=\partial_\Pi\log Z_{\Pi}=\partial_\Pi \log \int [d\bm{\Gamma}]\P(\A,\bm{\Gamma}|\Pi)=\frac{\int [d\bm{\Gamma}]\partial_\Pi\P(\A,\bm{\Gamma}|\Pi)}{\int [d\bm{\Gamma}]\P(\A,\bm{\Gamma}|\Pi)}=0.
\end{equation}
Since maximizing the likelihood in Eq. \ref{postParameters}, i.e. solving Eq. \ref{Eqlearn}, needs computing expectations with respect to the posterior in Eq. \ref{postHidden}, this is an Expectation-Maximization (EM) method \cite{friedmanLearningBook}. 
\subsection{Inference of time-varying parameters}\label{dynamicRAS} Let us assume to know the static parameters $\Pi$. We do not solve the inference problem for the time series of dynamic parameters $\ThetaB$ overall, i.e. by maximizing Eq. \ref{postHidden}. Instead, we infer step by step the parameters $\Theta^t$ by conditioning on the expectations $\hat{\Theta}^{t-1}$, that are the one step backward estimates\footnote{Here, we are assuming to know the expectation for $\Theta^0$, i.e. $\hat{\Theta}^0$. Below, we explain how to infer the initial point for the latent dynamics.} for $\Theta^{t-1}$. 

Let us focus on the inference at the generic time $t\neq 0$ when the previous network snapshot is observed and let $\mathcal{F}^t\equiv\{A^{t-1},\Pi\}$ be the information set for the considered problem. From the Bayes' theorem, it is
\begin{equation}\label{postTheta}
\P(\Theta^t|\A^t,\Theta^{t-1},\mathcal{F}^t)=\frac{\P(\A^t|\Theta^t,\mathcal{F}^t)\P(\Theta^t|\Theta^{t-1},\mathcal{F}^t)\P(\Theta^{t-1}|\mathcal{F}^t)}{\P(\A^{t},\Theta^{t-1}|\mathcal{F}^t)}.
\end{equation}
Hence, by conditioning on the expectation for $\Theta^{t-1}$, i.e. $\hat{\Theta}^{t-1}$, the inference problem can be solved by maximizing the following likelihood for $\Theta^t$,
\begin{equation}\label{fit1}
\P(\Theta^t|\bm{A}^t,\A^{t-1},\hat{\Theta}^{t-1},\Pi)\propto\P(\bm{A}^t|\A^{t-1},\Theta^t,\bm{\alpha})\P(\Theta^t|\hat{\Theta}^{t-1},\bm{\Phi})\:,\:\:\:\:\forall t=1,...,T.
\end{equation}
Maximizing Eq. \ref{fit1} is equivalent to solve the following problem
\begin{equation}\label{argmaxtheta}
\hat{\Theta}^{t}=\underset{{\gamma}}{\mbox{argmax}}\left(\log\P(\A^t|\A^{t-1},\gamma,\bm{\alpha})+\log F(\gamma|\hat{\Theta}^{t-1},\bm{\Phi})\right),
\end{equation}
where $F(\Theta^t|\Theta^{t-1},\bm{\Phi})\equiv\prod_{i=1}^N f(\theta_i^t|\phi_{0,i}+\phi_{1,i}\hat{\theta}_i^{t-1},\sigma_i^2)$ is the Gaussian probability density function associated with the transition probability for the latent variables. Eq. \ref{argmaxtheta} is equivalent to the following system of non linear equations,
\begin{equation}\label{eqTheta}
\left[\sum_{j\neq i}\left(\frac{(1-\alpha_{ij})\frac{e^{A_{ij}^t(\theta_i^t+\theta_j^t)}}{1+e^{(\theta_i^t+\theta_j^t)}}}{\alpha_{ij}\mathbb{I}_{A_{ij}^tA_{ij}^{t-1}}+(1-\alpha_{ij})\frac{e^{A_{ij}^t(\theta_i^t+\theta_j^t)}}{1+e^{(\theta_i^t+\theta_j^t)}}}\right)\left(-A_{ij}^t+\frac{e^{(\theta_i^t+\theta_j^t)}}{1+e^{(\theta_i^t+\theta_j^t)}}\right)\right]-\frac{\theta_i^t-\phi_{0,i}-\phi_{1,i}\hat{\theta}_{i}^{t-1}}{\sigma_i^2}=0,\:\:\:\forall i=1,...,N.
\end{equation}
This system can be solved by the following iterative proportional fitting procedure: (i) assume any starting point $\hat{\theta}_i^t$ $\forall i=1,...,N$ for the node fitness\footnote{A possible choice is $\hat{\theta}_i^t=\phi_{0,i}+\phi_{1,i}\hat{\theta}_{i}^{t-1}$}; (ii) then, solve one by one the equations in Eq. \ref{eqTheta} by conditioning on $\hat{\theta}_j^t$ $\forall j\neq i$; (iii) update the value for $\hat{\theta}_i^t$ with the solution of the corresponding equation; (iv) repeat until convergence.

The proposed method for the inference of the time-varying parameters is a statistical filtering algorithm. Filtering is an operation that involves the extraction of information about a latent quantity of interest at time $t$ by using data measured up to and including $t$ \cite{chen2003}, like in Kalman filter and its extensions. However, differently from Kalman filter, we study the case of a Hidden Markov Model with continuous-valued state space, i.e. continuous-valued state vector $\Theta^t$, but binary measurement matrix, i.e. $\A^t$. Finally, let us notice that the proposed method can be used for on-line inference: once the off-line learning of the static parameters is completed, we can solve the filtering problem for $\Theta^t$ in Eq. \ref{argmaxtheta} whenever the new measurement $\A^t$ is available. On-line inference is particularly useful for link prediction: let $\A^t$ be the observation at the current time and we want to construct the one-step-ahead forecast, i.e. $\E[\A^{t+1}|\A^t]$. Once $\hat{\Theta}^t$ is inferred on-line by solving Eq. \ref{argmaxtheta}, the one-step-ahead forecast is constructed by averaging over the probability distribution
$$
\P(\A^{t+1}|\A^t,\hat{\Theta}^t,\Pi)=\int [d\Theta^{t+1}]\P(\A^{t+1}|\A^{t},\Theta^{t+1},\bm{\alpha})\P(\Theta^{t+1}|\hat{\Theta}^{t},\bm{\Phi})
$$
obtained by projecting the latent state. In Section \ref{empirical} we show an application of this procedure.

\subsection{Learning $\bm{\alpha}$} Let assume to know the static parameters $\bm{\Phi}$ and consider the problem of learning $\bm{\alpha}$. The most likely estimate for $\bm{\alpha}$ is obtained by maximizing the associated posterior in Eq. \ref{postParameters}, that is
\begin{equation}\label{posteriorAlpha}
\P(\bm{\alpha}|\A)\propto\int \left[d\bm{\Theta}\right]\P(\A,\bm{\Theta}|\bm{\alpha},\bm{\Phi})=\int \prod_{t=1}^T\left[\:d\Theta^t\right]\P(\A^t|\A^{t-1},\Theta^t,\bm{\alpha})F(\Theta^t|\Theta^{t-1},\bm{\Phi}).
\end{equation}
However, the integral in Eq. \ref{posteriorAlpha} is infeasible because of the nonlinearity of the probability distribution of the network. Hence, we maximize an approximated likelihood where the transition probabilities for the latent variables are conditioned on the expectations at the previous step, i.e.
\begin{equation}\label{Zalpha}
l_{\bm{\alpha}}\equiv\int \prod_{t=1}^T\left[\:d\Theta^t\right]\P(\A^t|\A^{t-1},\Theta^t,\bm{\alpha})F(\Theta^t|\Theta^{t-1},\bm{\Phi})\approx\prod_{t=1}^T\int\left[d\Theta^t\right]\P(\A^t|\A^{t-1},\Theta^t,\bm{\alpha})F(\Theta^t|\hat{\Theta}^{t-1},\bm{\Phi})\equiv\tilde{l}_{\bm{\alpha}},
\end{equation}
where $\hat{\theta}_i^{t-1}$ is the expectation of $\theta_i^{t-1}$ that we obtained by solving Eq. \ref{eqTheta}.

Let us focus on the learning of parameter $\alpha_{ij}$. When we aim to obtain the solution for $\alpha_{ij}$, the only time-varying parameters that are involved in the learning are the ones associated with node $i$ and node $j$, i.e. $\{\theta_i^t\}^{t=1,...,T}$ and $\{\theta_j^t\}^{t=1,...,T}$. Hence, the most likely estimate for $\alpha_{ij}$ is the value that maximizes the following log-likelihood
\begin{equation}\label{Saij}
\tilde{S}_{\alpha_{ij}}=\log \tilde{l}_{\alpha_{ij}}=\sum_{t=1}^T\log\int dx dy \left(\alpha_{ij}\mathbb{I}_{A_{ij}^tA_{ij}^{t-1}}+(1-\alpha_{ij})\frac{e^{A_{ij}^t(x+y)}}{1+e^{(x+y)}}\right)f(x|\phi_{0,i}+\phi_{1,i}\hat{\theta}_i^{t-1},\sigma_i^2)f(y|\phi_{0,j}+\phi_{1,j}\hat{\theta}_j^{t-1},\sigma_j^2).
\end{equation}

In the learning procedure, the following double integral is involved,
\begin{eqnarray}\label{integralLearningAlpha}
&&\mathcal{I}_{A_{ij}^t}(\hat{\theta}_i^{t-1},\hat{\theta}_j^{t-1},\bm{\Phi}_i,\bm{\Phi}_j)=\\ &&
\int dxdy\frac{e^{A_{ij}^t(x+y)}}{1+e^{(x+y)}}f(x|\phi_{0,i}+\phi_{1,i}\hat{\theta}_i^{t-1},\sigma_i^2)f(y|\phi_{0,j}+\phi_{1,j}\hat{\theta}_j^{t-1},\sigma_j^2), \nonumber
\end{eqnarray}
which can be solved numerically. However, we propose to apply the following integral identity proposed by Polson et al. \cite{polson2013} 
\begin{equation}\label{polson}
\frac{(e^\psi)^a}{(1+e^\psi)^b}=2^{-b}e^{(a-\frac{b}{2})\psi}\int_0^\infty e^{-\frac{\omega\psi^2}{2}}p_{PG}(\omega)d\omega
\end{equation}
where $b>0$, $a,\psi\in\mathbb{R}$, and $p_{PG}:[0,\infty)\mapsto[0,1)$ is the density of the P\'{o}lya-Gamma distribution. There is no a closed-form expression for $p_{PG}$ but we evaluate it numerically. A method for sampling P\'{o}lya-Gamma random variates can be found in \cite{polson2014}. The double integral in Eq. \ref{integralLearningAlpha} is equivalent to the following integral,
\begin{equation}\label{IAij}
\mathcal{I}_{A_{ij}^t}(\hat{\theta}_i^{t-1},\hat{\theta}_j^{t-1},\bm{\Phi}_i,\bm{\Phi}_j)\equiv\int_0^{\infty}\frac{d\omega}{2}p_{PG}(\omega)\mathcal{K}_{A_{ij}^t}(\omega,\hat{\theta}_i^{t-1},\hat{\theta}_j^{t-1},\bm{\Phi}_i,\bm{\Phi}_j)
\end{equation}
where
$$
\mathcal{K}_{A_{ij}^t}(\omega,\hat{\theta}_i^{t-1},\hat{\theta}_j^{t-1},\bm{\Phi}_i,\bm{\Phi}_j)=\frac{exp\left(\frac{\sigma_i^2+\sigma_j^2+4(\phi_{0,i}+\phi_{1,i}\hat{\theta}_i^{t-1}+\phi_{0,j}+\phi_{1,j}\hat{\theta}_j^{t-1})(2A_{ij}^t-1-\omega(\phi_{0,i}+\phi_{1,i}\hat{\theta}_i^{t-1}+\phi_{0,j}+\phi_{1,j}\hat{\theta}_j^{t-1}))}{8(1+\omega(\sigma_i^2+\sigma_j^2))}\right)}{\sqrt{1+\omega(\sigma_i^2+\sigma_j^2)}}.
$$
We propose to evaluate numerically the integral in Eq. \ref{IAij}. This gives the advantage of computing a single integral.

Then $\alpha_{ij}$ is estimated by solving the equation $\partial_{\alpha_{ij}} \tilde{S}_{\alpha_{ij}}=0$, which can be explicitly rewritten as
\begin{equation}\label{EqAij}
\sum_{t=1}^T\frac{\mathbb{I}_{A_{ij}^tA_{ij}^{t-1}}-\mathcal{I}_{A_{ij}^t}(\hat{\theta}_i^{t-1},\hat{\theta}_j^{t-1},\bm{\Phi}_i,\bm{\Phi}_j)}{\alpha_{ij}\mathbb{I}_{A_{ij}^tA_{ij}^{t-1}}+(1-\alpha_{ij})\mathcal{I}_{A_{ij}^t}(\hat{\theta}_i^{t-1},\hat{\theta}_j^{t-1},\bm{\Phi}_i,\bm{\Phi}_j)}=0.
\end{equation}
The solution of Eq. \ref{EqAij} represents the most likely estimate $\hat{\alpha}_{ij}$ given the data.

\subsection{Learning $\bm{\Phi}$}  Let assume to know the static parameters $\bm{\alpha}$ and consider the problem of learning $\bm{\Phi}$. Similarly to the previous Subsection, we use conditions on the expectations for the latent variables to obtain an approximated log-likelihood for $\bm{\Phi}$,
\begin{equation}\label{SPHI}
\tilde{S}_{\bm{\Phi}}=\sum_{t=1}^T\log\int\left[\prod_{k=1}^Ndx_k\:f(x_k|\phi_{0,k}+\phi_{1,k}\hat{\theta}_k^{t-1},\sigma_k^2)\right]\left[\prod_{i,j>i}\alpha_{ij}\mathbb{I}_{A_{ij}^tA_{ij}^{t-1}}+(1-\alpha_{ij})\frac{e^{A_{ij}^t(x_i+x_j)}}{1+e^{(x_i+x_j)}}\right].
\end{equation}

Let us focus on the learning of parameters $\bm{\Phi}_i\equiv\{\phi_{0,i},\phi_{1,i},\sigma_i\}$. Because of the marginal distribution, each time-varying parameter $\theta_i^t$ is coupled with all the others and this prevents the valuation of the multiple integral in Eq. \ref{SPHI}. Hence, we adopt the following approximation for the probability measure,
$$
\prod_{k=1}^Ndx_k\:f(x_k|\phi_{0,k}+\phi_{1,k}\hat{\theta}_k^{t-1},\sigma_k^2)\approx dx_i\:f(x_i|\phi_{0,i}+\phi_{1,i}\hat{\theta}_i^{t-1},\sigma_i^2)\prod_{k\neq i}dx_k\:\delta(x_k-\hat{\theta}_k^t)f(x_k|\phi_{0,k}+\phi_{1,k}\hat{\theta}_k^{t-1},\sigma_k^2),
$$
i.e. we condition on the expectations at time $t$ for all the latent variables with the exception of $\theta_i^t$. Then we maximize the following quantity,
\begin{equation}\label{SPHIi}
\tilde{S}_{\bm{\Phi}_i}=\sum_{t=1}^T\log\int_{-\infty}^{\infty} dx_i\:f(x_i|\phi_{0,i}+\phi_{1,i}\hat{\theta}_i^{t-1},\sigma_i^2)\left(\prod_{j\neq i}\alpha_{ij}\mathbb{I}_{A_{ij}^tA_{ij}^{t-1}}+(1-\alpha_{ij})\frac{e^{A_{ij}^t(x_i+\hat{\theta}_j^t)}}{1+e^{(x_i+\hat{\theta}_j^t)}}\right),
\end{equation}
i.e. we estimate $\bm{\Phi}_i$ by solving the system of equations $\partial_{\bm{\Phi}_i}\tilde{S}_{\bm{\Phi}_i}=0$. Let us define the following partition function $\forall t=1,...,T$,
\begin{equation}\label{Pit}
Z_{\bm{\Phi}_i}^t\equiv\int_{-\infty}^{+\infty}dx\: f(x_i|\phi_{0,i}+\phi_{1,i}\hat{\theta}_i^{t-1},\sigma_i^2)\left(\prod_{j\neq i}\alpha_{ij}\mathbb{I}_{A_{ij}^tA_{ij}^{t-1}}+(1-\alpha_{ij})\frac{e^{A_{ij}^t(x+\hat{\theta}_j^t)}}{1+e^{(x+\hat{\theta}_j^t)}}\right)
\end{equation}
and let $\mu_{\bm{\Phi}_i}^t$ and $\Sigma_{\bm{\Phi}_i}^t$ be the first and the second moment of the distribution, respectively.

The system of equations $\partial_{\bm{\Phi}_i}\tilde{S}_{\bm{\Phi}_i}=0$ reads explicitly as
\begin{equation}\label{eqphiexp}
\begin{cases}
\langle\bm{\mu}_{\bm{\Phi}_i}\rangle-\phi_{0,i}-\langle L\hat{\bm{\theta}}_{i}\rangle\phi_{1,i} &= 0\\
\frac{1}{T} (L\hat{\bm{\theta}}_{i}^\intercal\:\bm{\mu}_{\bm{\Phi}_i})-\langle L\hat{\bm{\theta}}_{i}\rangle\phi_{0,i}-\frac{1}{T} (L\hat{\bm{\theta}}_{i}^\intercal\: L\hat{\bm{\theta}}_{i})\phi_{1,i} &= 0\\
\sigma_i^2-\left(\langle \bm{\Sigma}_{\bm{\Phi}_i}\rangle+\phi_{0,i}^2+\frac{1}{T} (L\hat{\bm{\theta}}_{i}^\intercal\:L\hat{\bm{\theta}}_{i})\phi_{1,i}^2-2\langle\bm{\mu}_{\bm{\Phi}_i}\rangle\phi_{0,i}-2\frac{1}{T} (L\hat{\bm{\theta}}_{i}^\intercal\:\bm{\mu}_{\bm{\Phi}_i})\phi_{1,i}+2\langle L\hat{\bm{\theta}}_{i}\rangle\phi_{0,i}\phi_{1,i}\right) &=0
\end{cases}
\end{equation}
where bold symbols represent $T$-dimensional vectors, i.e. $\bm{x}=(x^1,x^2,...,x^T)'$, angle brackets denote time average, i.e. $\langle\bm{x}\rangle\equiv\frac{1}{T}\sum_{t=1}^Tx^t$, and $L$ is the \emph{lag operator}, i.e. $Lx^t=x^{t-1}$. Let us notice that $L\hat{\theta}_i^1=\hat{\theta}_i^0$ represents the latent state at the initial time (see below).

The system of nonlinear equations can be solved with the following iterative proportional fitting procedure: (i) assume any starting point $\bm{\Phi}_i^0$; (ii) compute $\mu^t_{\bm{\Phi}_i^0}$ and $\Sigma^t_{\bm{\Phi}_i^0}$ $\forall t=1,...,T$; (iii) solve the system of equation in Eq. \ref{eqphiexp} by substituting $\mu^t_{\bm{\Phi}_i^0}\rightarrow\mu^t_{\bm{\Phi}_i}$ and $\Sigma^t_{\bm{\Phi}_i^0}\rightarrow\Sigma^t_{\bm{\Phi}_i}$ $\forall t=1,...,T$; (iv) update the values for $\bm{\Phi}_i^0$ and continue until convergence.

\subsection{The algorithm} The estimation procedure consists in alternating the inference of the latent variables in the Expectation step and the learning of the static parameters in the Maximization step until convergence. As a starting point of the method, the time-varying parameters $\{\Theta^t\}^{t=0,1,...,T}$ can be estimated by single snapshot inference, i.e. given the network snapshot at time $t$ and by assuming $\P(\A^t|\bm{\gamma})=\prod_{i,j>i}\frac{e^{A_{ij}^t(\gamma_i+\gamma_j)}}{1+e^{(\gamma_i+\gamma_j)}}$, we solve snapshot by snapshot the following problem,
\begin{equation}\label{staticProblem}
\tilde{\Theta}^{t}=\underset{{\gamma}}{\mbox{argmax}}\:\:\log\P(\A^t|\gamma)\:\:\:\:\forall t=0,1,...T
\end{equation}
and we obtain a naive estimation $\{\tilde{\Theta}^t\}^{t=0,1,...,T}$ of the hidden states of the Markov chain in Eq. \ref{dartgrg}. In particular, we infer the latent state at the initial time, i.e. $\hat{\Theta}^0\equiv\tilde{\Theta}^0$.

Then, we estimate the process in Eq. \ref{ar1} for the naively inferred $\{\tilde{\Theta}^t\}^{t=0,1,...,T}$ to obtain a naive estimation of the static parameters $\tilde{\bm{\Phi}}$. Finally, the naive estimate $\tilde{\bm{\alpha}}$ for the probabilities of copying can be obtained by solving Eq. \ref{EqAij} with naively inferred $\{\tilde{\Theta}^t\}^{t=0,1,...,T}$ and $\tilde{\bm{\Phi}}$. We refer to this naive estimation method as the Single Snapshot Inference (SSI) of the model.

Hence, we apply the following iterative algorithm:
\begin{enumerate}
\item Assume as starting point $\tilde{\ThetaB}$ and $\tilde{\Pi}=\{\tilde{\bm{\Phi}},\tilde{\bm{\alpha}}\}$.
\item Infer $\hat{\bm{\Theta}}\equiv\{\hat{\Theta}^t\}^{t=1,...T}$ by solving Eq. \ref{eqTheta} with $\tilde{\Pi}$.
\item Learn $\hat{\bm{\alpha}}$ by solving Eq. \ref{EqAij} for each possible couple of nodes with previously inferred $\hat{\ThetaB}$ and $\tilde{\bm{\Phi}}$.
\item Learn $\hat{\bm{\Phi}}$ by solving Eq. \ref{eqphiexp} for each $i$ with previously inferred $\hat{\ThetaB}$ and $\hat{\bm{\alpha}}$.
\item Update $\tilde{\ThetaB}\leftarrow\hat{\ThetaB}$.
\item Update $\tilde{\Pi}\leftarrow\hat{\Pi}$.
\item Repeat until convergence.
\end{enumerate}
This is an Expectation-Maximization learning algorithm \cite{dempster1977}, where we use a generalization of the RAS algorithm \cite{bacharach1965} for the expectation step (line 2). The RAS algorithm is usually adopted to solve the problem of estimating nonnegative matrices from marginal data\footnote{The problem in Eq. \ref{staticProblem} can be solved with the RAS algorithm where the generic entry of the matrix is $\frac{e^{\gamma_i+\gamma_j}}{1+e^{\gamma_i+\gamma_j}}$ and the marginal data are represented by the degree sequence.} and is preferred to other methods due to its computational speed, numerical stability and algebraic simplicity. In Subsection \ref{dynamicRAS} we generalize the RAS algorithm to the case of time-varying parameters. The main cycle of the algorithm takes $O(N\times T)$ time. The number of iterations needed for the generalized RAS algorithm to converge is not deterministic similarly to the original one. However, we observe numerically it takes $O(10^{0\div 1})$ iterations when $N$ is $O(10^{2\div 3})$. The number of operations needed for the maximization step (lines 3 and 4) is, in general, a more complicated question. Learning $\bm{\alpha}$ takes $O(N^2)$ steps, one for each $\alpha_{ij}$, and each step takes $T+1$ operations, the numerical evaluation of $T$ single integrals and finding the zero of a function. Learning $\bm{\Phi}$ takes $N$ steps, one for each $\bm{\Phi}_i$, but each step takes a non deterministic number of cycles in order to solve the system of integral equations in Eq. \ref{eqphiexp}. In average, each step takes $O(10^{1\div 2})$ cycles when $T$ is $O(10^2)$. Each cycle takes $3\times T$ operations, i.e. the numerical evaluation of $3\times T$ single integrals. Finally, the number of iterations for the EM algorithm to converge is not deterministic but we observe numerically that it is quite constant in the size of the system.
\begin{table}
\begin{tabular}{SSSSS} \toprule
    {} & {$N=100$} & {$N=250$} & {$N=500$} & {$N=1000$} \\ \midrule
   {time ($h$)}   & 2.8(5) & 11(1) & 45(4) & 151(12)  \\ \bottomrule \vspace{0.1cm}
\end{tabular}
\caption{The average time of convergence for the EM algorithm applied to the DAR-TGRG model in the case of undirected networks with $T=200$. In the model simulations, the parameters are randomly determined as explained in Section \ref{montecarlo}. The simulations were performed using a Matlab code executed on an ordinary dual-core Intel Core i5, with 8 GB RAM.}
\label{tab0}
\end{table}
Table \ref{tab0} shows how much time the EM algorithm takes in average to converge.

\section{Monte Carlo simulations}\label{montecarlo}
Before applying our methodology to real data, we run Monte Carlo simulations to study the performance of the proposed estimation method when applied both to undirected and to directed networks. Data are generated according to the described models with randomly chosen static parameters. In the case of undirected networks, the DAR(1) model parameters are sampled uniformly in the unit interval. For TGRG we sample $\phi_{1,i}\sim U(-1,1)$, $\sigma_i\sim U(0,1)$ and $\phi_{0,i}\sim {\mathcal N}(0,1)$.  For DAR-TGRG, $\alpha_{ij}\sim U(0,1)$. For both models time-varying parameter $\theta_i^t$ follows the stationary AR(1) process of Eq. \ref{ar1}. We estimate the models with the proposed Expectation-Maximization (EM) algorithm and compare the results also with the Single Snapshot Inference (SSI). For each simulation, we estimate $\bm{\Phi}_i$ for each node $i$. For DAR-TGRG model we obtain also $\binom{N}{2}$ estimates for $\alpha_{ij}$, one for each possible couple of nodes $(i,j)$. For both models, we infer the time series of the latent variables $\{\theta_i^t\}_{i=1,...,N}^{t=0,1,...,T}$. We simulate each model $100$ times. In evaluating the goodness of fit of the proposed estimation method, we report the mean absolute relative error for the estimate of parameters. The mean is obtained by averaging over the nodes and the number of simulations. For the time-varying parameters we consider also the time average of the absolute relative errors. A similar study is performed for the case of directed networks, with the exception that for each node we have two fitnesses, $\theta_i^{t,out}$ and $\theta_i^{t,in}$, and as a consequence two sets of static parameters $\bm{\Phi}\equiv\{\bm{\Phi^{out}},\bm{\Phi^{in}}\}$. For DAR-TGRG model, we obtain $N(N-1)$ estimates for $\alpha_{ij}$, one for each possible couple of ordered nodes.

\begin{table}
\begin{tabular}{SSSSS} \toprule
    {} & {$T=100$} & {$T=200$} & {$T=500$} & {$T=1000$} \\ \midrule
   $\chi_{ij}$   & 0.28 & 0.21 & 0.16 & 0.10  \\ 
   $\alpha_{ij}$   & 0.29 & 0.19 & 0.12 & 0.08 \\ \bottomrule \vspace{0.1cm}
\end{tabular}
\caption{The mean absolute relative error of the estimates of parameters of the DAR(1) model as a function of the length $T$ of time series. We simulate the DAR(1) model $100$ times.}
\label{tab1}
\end{table}

\begin{table}
\begin{tabular}{SSSSSS} \toprule
   {} & {$N$} & {$\theta_i^t$} & {$\phi_{0,i}$} & {$\phi_{1,i}$} & {$\sigma_i$} \\ \midrule
{SSI} &  100  & 0.30 & 0.58 & 0.46 & 0.69 \\
 {EM} &  100  & 0.22 & 0.13 & 0.13 & 0.06 \\ \midrule
 {SSI} & 200  & 0.20 & 0.31 & 0.27 & 0.31  \\
  {EM} & 200  & 0.10 & 0.10 & 0.10 & 0.05  \\  \bottomrule \vspace{0.1cm}
\end{tabular}
\caption{The mean absolute relative error of the estimates of parameters for the TGRG model in the case of undirected networks. We set $T=200$ and simulate the model $100$ times. We compare the proposed Expectation-Maximization algorithm (EM) with the Single Snapshot Inference (SSI).}
\label{tab2}
\end{table}
\begin{table}
\begin{tabular}{SSSSSS} \toprule
   {} & {$N$} & {$\theta_i^{t,out(in)}$} & {$\phi_{0,i}^{out(in)}$} & {$\phi^{out(in)}_{1,i}$} & {$\sigma^{out(in)}_i$} \\ \midrule
{SSI} &  100  & 0.31 & 0.59 & 0.47 & 0.71 \\
 {EM} &  100  & 0.23 & 0.12 & 0.12 & 0.06 \\ \midrule
 {SSI} & 200  & 0.21 & 0.33 & 0.29 & 0.33  \\
  {EM} & 200  & 0.10 & 0.11 & 0.10 & 0.05  \\  \bottomrule \vspace{0.1cm}
\end{tabular}
\caption{The mean absolute relative error of the estimates of parameters for the TGRG model in the case of directed networks. We set $T=200$ and simulate the model $100$ times. We compare the proposed Expectation-Maximization algorithm (EM) with the Single Snapshot Inference (SSI).}
\label{tab2bis}
\end{table}
\begin{table}
\begin{tabular}{SSSSSS} \toprule
  {} &  {$\alpha_{ij}$} & {$\theta_i^t$} & {$\phi_{0,i}$} & {$\phi_{1,i}$} & {$\sigma_i$} \\ \midrule
 {SSI} &  0.22  & 0.29 & 0.27 & 0.18 & 0.22  \\
 {EM} &  0.18  & 0.14 & 0.15 & 0.10 & 0.06  \\ \bottomrule \vspace{0.1cm}
\end{tabular}
\caption{The mean absolute relative error of the estimates of parameters for the DAR-TGRG model in the case of undirected networks. We compare the EM algorithm with the single snapshot inference SSI. We set $N=200$, $T=200$ and simulate the model $100$ times.}
\label{tab3}
\end{table}
\begin{table}
\begin{tabular}{SSSSSS} \toprule
  {} &  {$\alpha_{ij}$} & {$\theta_i^{t,out(in)}$} & {$\phi_{0,i}^{out(in)}$} & {$\phi_{1,i}^{out(in)}$} & {$\sigma_i^{out(in)}$} \\ \midrule
 {SSI} &  0.22  & 0.30 & 0.28 & 0.18 & 0.23  \\
 {EM} &  0.17  & 0.14 & 0.14 & 0.10 & 0.05  \\ \bottomrule \vspace{0.1cm}
\end{tabular}
\caption{The mean absolute relative error of the estimates of parameters for the DAR-TGRG model in the case of directed networks. We compare the EM algorithm with the single snapshot inference SSI. We set $N=200$, $T=200$ and simulate the model $100$ times.}
\label{tab3bis}
\end{table}

The simulation results are summarized in Tables \ref{tab1}, \ref{tab2}, \ref{tab2bis}, \ref{tab3} and \ref{tab3bis}. In Table \ref{tab1} we show the results for the maximum likelihood estimation of the DAR(1) process in Eq. \ref{dar1}. The remaining Tables show that the EM method greatly outperforms the single snapshot inference SSI. Furthermore, we find that the mean absolute relative error for both EM and SSI declines with the number of nodes $N$ since the number of observations increases as $N^2$, while the number of parameters increases linearly with $N$. 
\begin{figure}[t]
\centering
{\includegraphics[width=\textwidth]{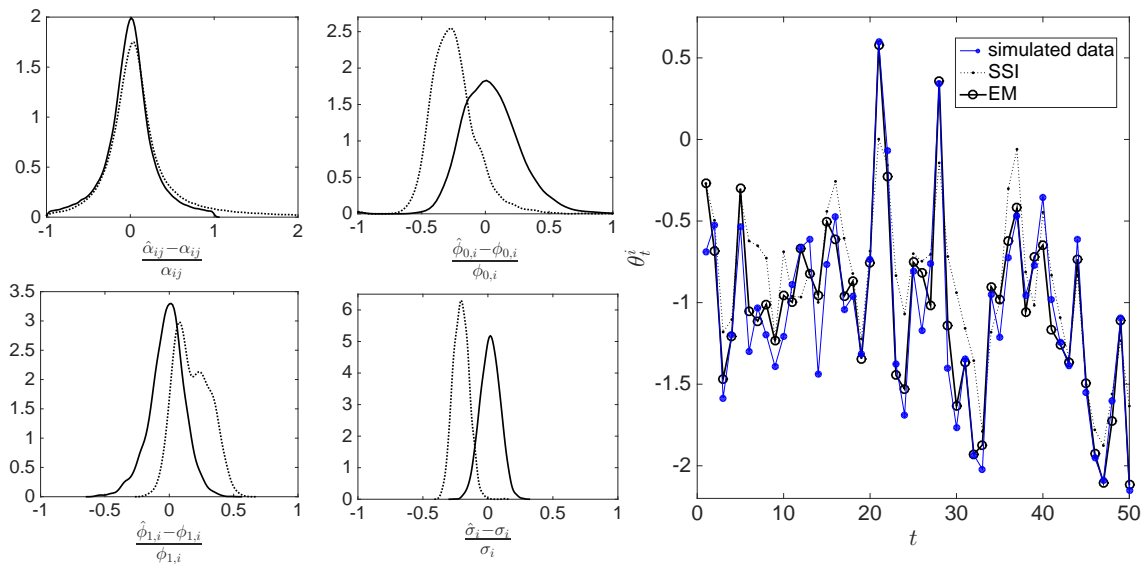}}
\caption{Left panels: estimated density of relative errors of $\alpha_{ij}$, $\phi_{0,i}$, $\phi_{1,i}$ and $\sigma_i$. We compare the EM method (solid line) with SSI (dotted line). Right panel: latent dynamics for a generic $\theta_{i}^t$ compared with the inferred one according to EM and SSI.}
\label{GOF}
\end{figure}
Figure \ref{GOF} shows the estimated density of the relative errors of the static parameters of the DAR-TGRG model in the case of undirected networks by applying both EM and SSI estimation methods. The SSI leads to biased estimation of the static parameters $\bm{\Phi}_i$, while the estimation obtained with the proposed EM method is unbiased. For illustrative purposes,  in the right panel of figure \ref{GOF} we show a typical sample realization and the estimate of time-varying parameter $\theta_i^t$. Clearly, the values inferred with EM track the simulated data fairly closely. Table \ref{tab4} shows how the mean absolute relative error of the parameters of the DAR-TGRG model of a dynamic undirected graph decreases with the length of the time series.

\begin{table}
\begin{tabular}{SSSS} \toprule
    {}  & {$T=300$} & {$T=500$} & {$T=1000$} \\ \midrule
   $\theta_{i}^{t}$    & 0.13 & 0.13 & 0.12  \\ 
   $\alpha_{ij}$    & 0.13 & 0.10 & 0.08  \\ 
   $\phi_{0,i}$    & 0.10 & 0.09 & 0.07  \\
   $\phi_{1,i}$    & 0.09 & 0.08 & 0.07  \\ 
   $\sigma_{i}$   & 0.05 & 0.04 & 0.04 \\ \bottomrule \vspace{0.1cm}
\end{tabular}
\caption{Mean absolute relative error of the parameters for the DAR-TGRG model estimated via the EM algorithm. The network is undirected. We set $N=200$ and the number of simulations is equal to $100$.}
\label{tab4}
\end{table}

\begin{figure}[t]
\centering
{\includegraphics[width=0.75\textwidth]{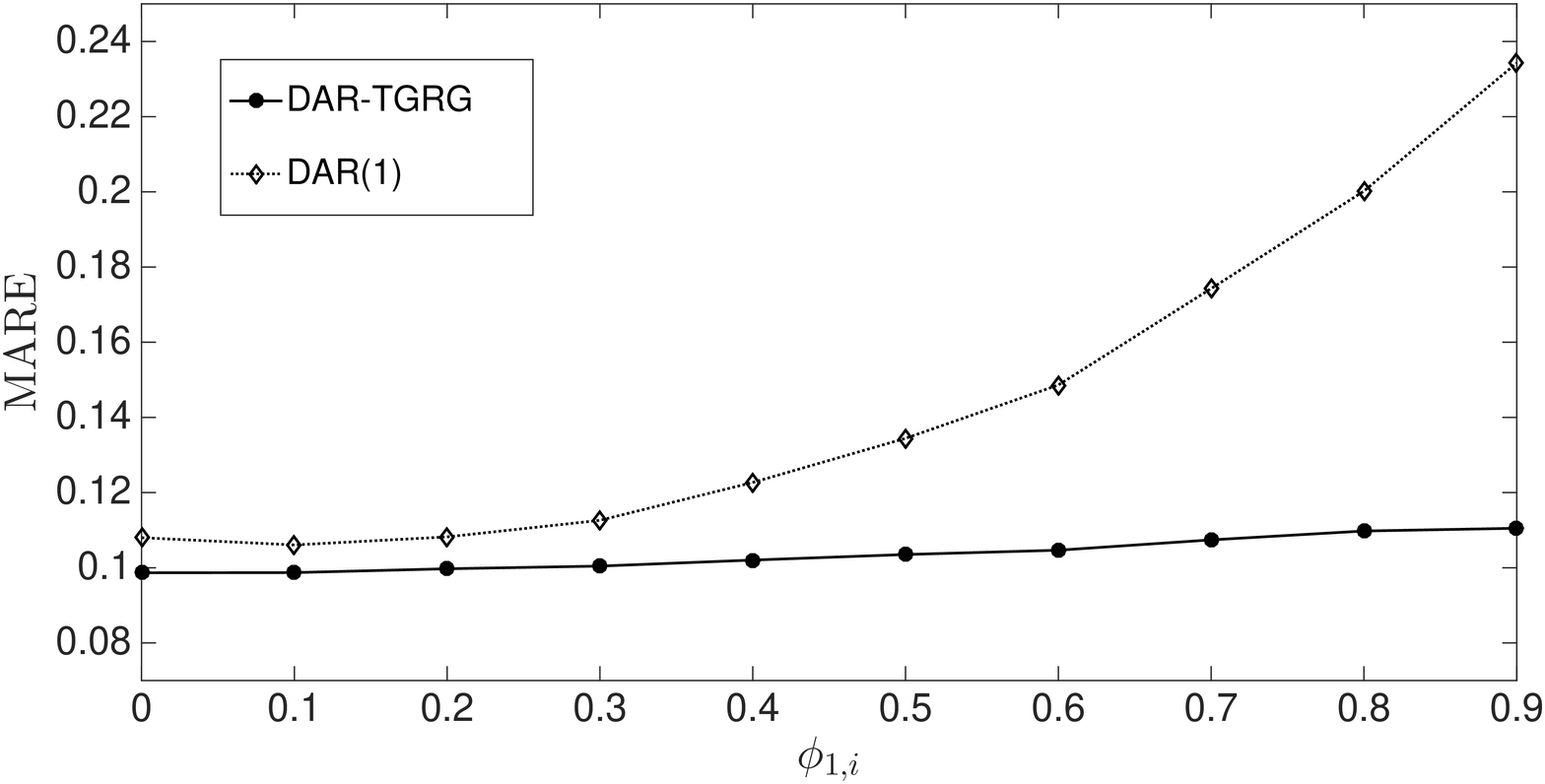}}
\caption{Mean Absolute Relative Error (MARE) of the estimates of $\alpha_{ij}$ as a function of the autoregressive coefficient $\phi_{1,i}$ for the time-varying parameters $\theta_i^t$. In the simulation of the DAR-TGRG model for the case of undirected networks, $\alpha_{ij}$, $\phi_{0,i}$ and $\sigma_i$ are randomly sampled while $\phi_{1,i}$ are equal for all $i$. We compare the goodness of fit of the estimates of $\alpha_{ij}$ via the EM method for the DAR-TGRG model (solid line) with the Maximum Likelihood (ML) estimates of $\alpha_{ij}$ according to the DAR(1) model (dotted line). We set $N=200$ and $T=400$.}
\label{AlphaEstimation}
\end{figure}

When the dynamics of the link is affected both by link persistence and by dynamic fitness, neglecting the last one can lead to an overestimation of the importance of the persistence. To show this we simulate a DAR-TGRG model for undirected networks taking $\phi_{1,i}$ equal for all time-varying parameters $\theta_i^t$\footnote{$\phi_{1,i}$ determines the autocorrelation of node fitness and as a consequence the link persistence associated with the time-varying marginal.}.  Then we estimate $\alpha_{ij}$ according to a DAR(1) model (which neglects dynamic fitness) and to a DAR-TGRG model.  Figure \ref{AlphaEstimation} shows the mean absolute error of $\alpha_{ij}$ for the two estimations as a function of  $\phi_{1,i}$. When this parameter is small both the DAR(1) model  and the DAR-TGRG model perform quite equivalently. On the contrary, when the dynamic fitness has a significant persistence due to a high value of $\phi_{1,i}$, the DAR(1) model wrongly imputes this to a link persistence which now has a large bias with respect to the DAR-TGRG model.

\section{Empirical application: understanding link persistence in the interbank market}\label{empirical}

Trading and credit networks are a natural application case for dynamic networks with persistence, like the one described by our model. Financial institutions lend mutually money on a daily basis and interbank markets are considered an important channel of propagation of systemic risk. While there is a vast literature on the static case, only few papers deal with the dynamic property of interbank networks. The static fitness model has been proved to characterize quantitatively several topological properties of the e-MID network \cite{demasi2006,musumeci2013}, to outperform other network models in the problem of reconstructing the e-MID network from limited information \cite{gabrielli2014,pm2017} and to give useful insights for systemic risk analysis of the interbank market \cite{cimini2015sr}. The ability of the fitness model to describe the static interbank network indicates that the size of two banks correlates with the existence of a credit between them. However it has been documented \cite{iorimarotta2015} the presence of memory effects in the process of network formation for interbank markets, according to the idea that a borrower, having asked for a loan many times to a lender in the past, is more likely to borrow from that lender again in the future than from other lenders, with which the borrower has never (or infrequently) interacted. 

In this section we estimate our dynamic model on data of an interbank market to disentangle the relative importance of fitness and link persistence in determining the future state of the network. This will allow also to perform a forecasting exercise to predict the existence of a credit relation between two banks.

\subsection{Data}

We investigate data from the electronic Market of Interbank Deposit (e-MID), a market where banks extend loans to one another for a specified term and/or collateral. A significant fraction of interbank loans are for maturities of one week or less, the majority being overnight. The e-MID is an electronic market in the Euro Area and it was founded in Italy in 1990 for Italian Lira transactions and denominated in Euros in 1999. According to the ``Euro Money Market Study 2006" published by the European Central Bank in February 2007, e-MID accounted for $17\%$ of total turnover in unsecured money market in the Euro Area. More recently the amount of overnight lending in e-MID has significantly declined, especially around the sovereign debt crisis \cite{baruccaemid1}. The e-MID network has been thoroughly studied to understand bank liquidity management, as for instance in \cite{baruccaemid1,ioriemidanalysis,frikeemidanalysis}.

The dataset contains the edge list of all credit transactions in each day from March $9^{th}$, 2012 to February $27^{th}$, 2015. In our analysis, we investigate the interbank network aggregated weekly. Each network snapshot of interbank deposits is constructed from the list of transactions where a bank, the lender, extends a loan to another bank, the borrower, that repays the loan in seven days, at most. Hence, we exclude loans with a term larger than a week. However, we account approximately for the $92\%$ of all the traded volume in the market since there are few credit relations with longer maturity. Then, we describe the e-MID weekly network with the unweighted and directed adjacency matrix $\A^t$: a generic element $A_{ij}^t$ is $1$ if the bank $i$ lends money at least once to bank $j$ during the week $t$, $0$ otherwise. We do not consider banks that interact less than $5\%$ of times in the considered period, i.e. in a period of $T=156$ weeks a bank has at least a credit relation for more than $7$ weeks. Hence, the credit network is formed by $N=98$ banks.

\subsection{Estimated fitness and link persistence in e-MID}

\begin{figure}[t]
\centering
{\includegraphics[width=0.49\textwidth]{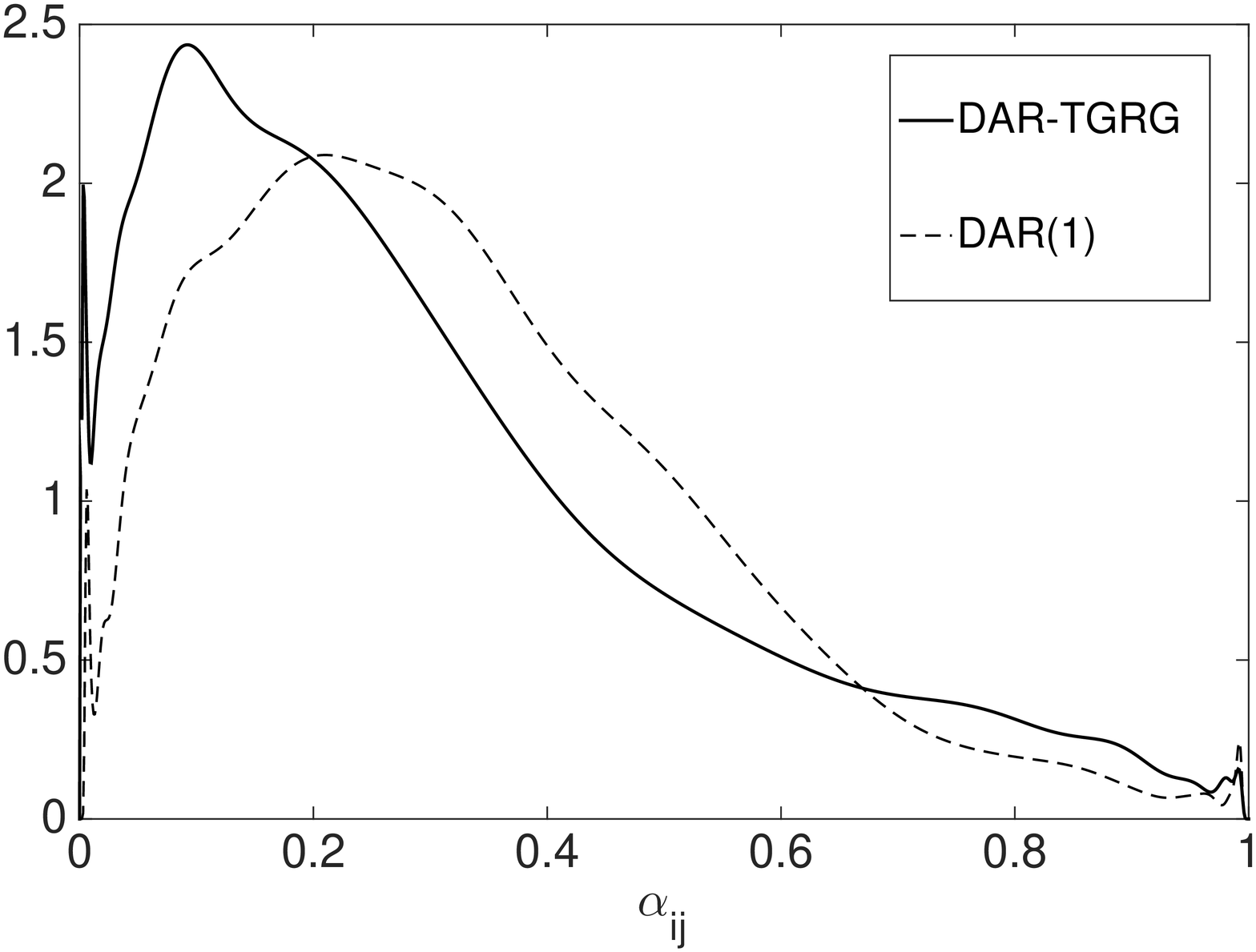}}
{\includegraphics[width=0.50\textwidth]{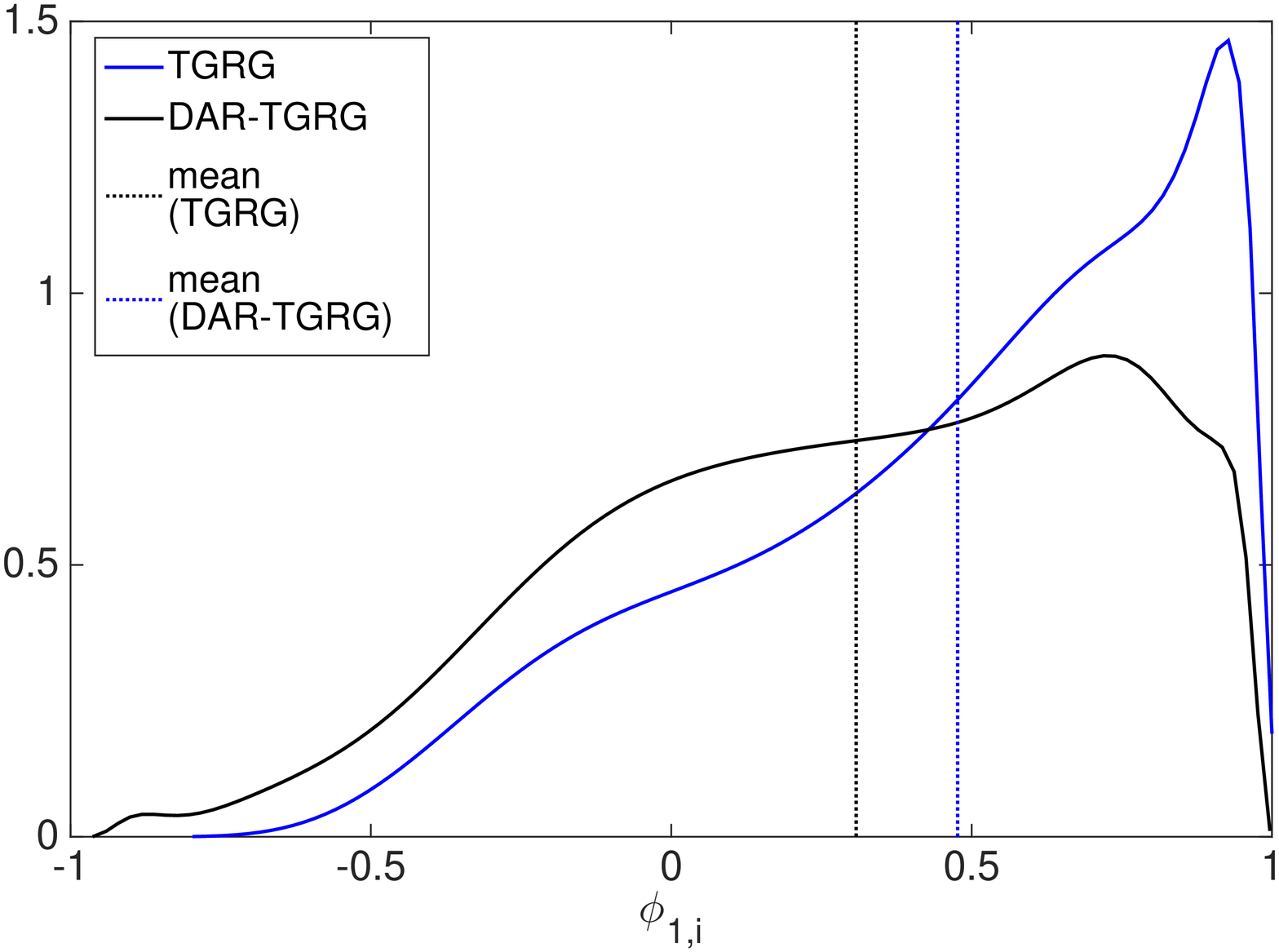}}
\caption{Left panel: distributions of $\alpha_{ij}$ estimated via EM on e-MID data. The solid line refers to the DAR-TGRG model while the dotted line refers to the DAR(1) model. Right panel: distribution of parameters $\phi_{1,i}$ estimated via EM. The black line refers to DAR-TGRG while the blue line to TGRG. The dotted lines represent the mean of the two distributions.}
\label{distpar}
\end{figure}

We estimate the three models on the time series of e-MID networks. Figure \ref{distpar} shows the estimated density of the $\alpha_{ij}$ link parameters (left panel) and of the $\phi_{1,j}$ node parameters (right panel) for the different model. We see that the DAR(1) model estimates larger $\alpha_{ij}$ parameters, i.e. larger link persistence, than the DAR-TGRG model. Similarly, the TGRG model estimates larger  $\phi_{1,j}$, i.e larger fitness persistence, than the complete DAR-TGRG model. Thus the full model balances the relative role of the two persistence mechanisms.

\begin{figure}[t]
\centering
{\includegraphics[width=0.49\textwidth]{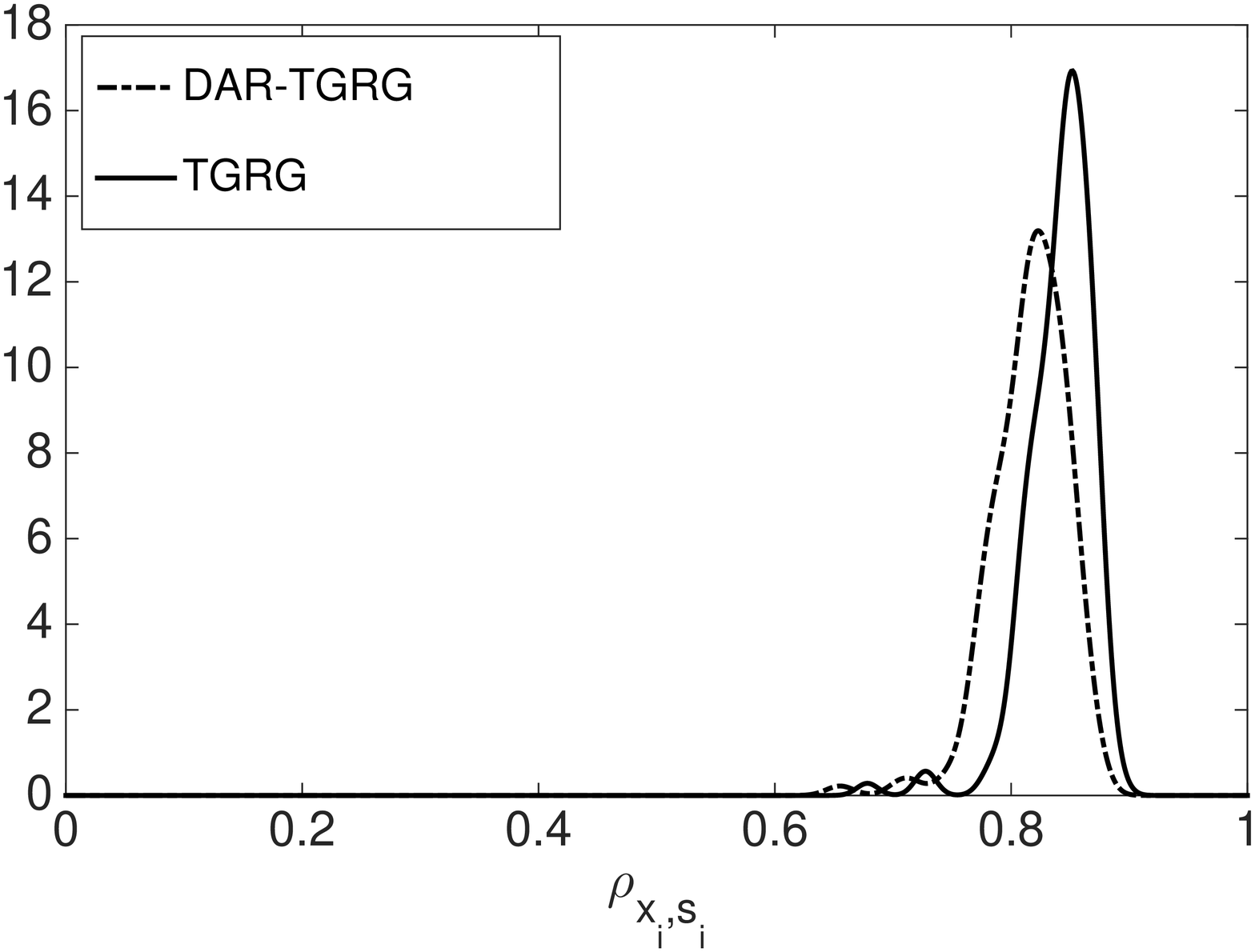}}
{\includegraphics[width=0.49\textwidth]{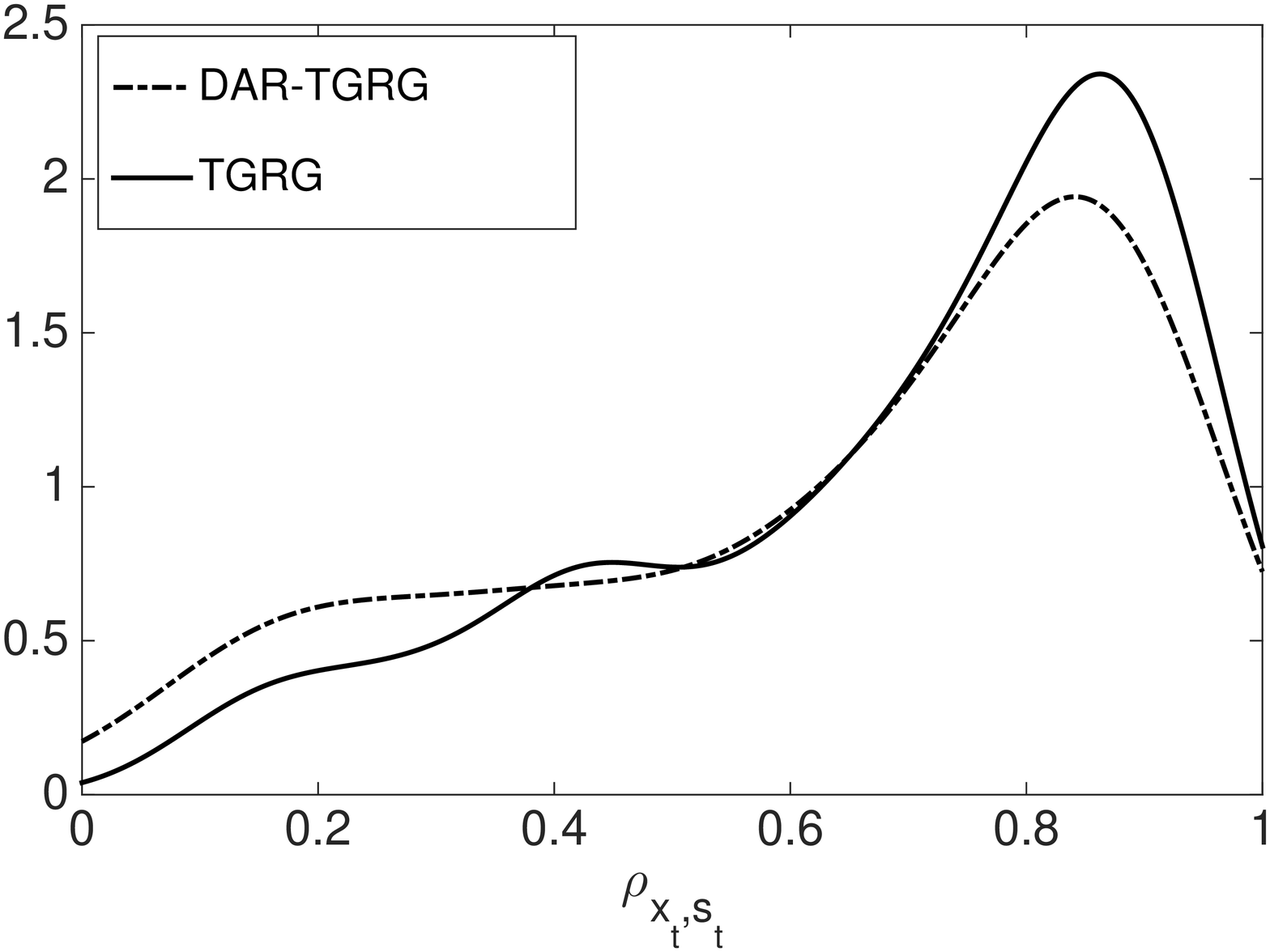}}
{\includegraphics[width=0.8\textwidth]{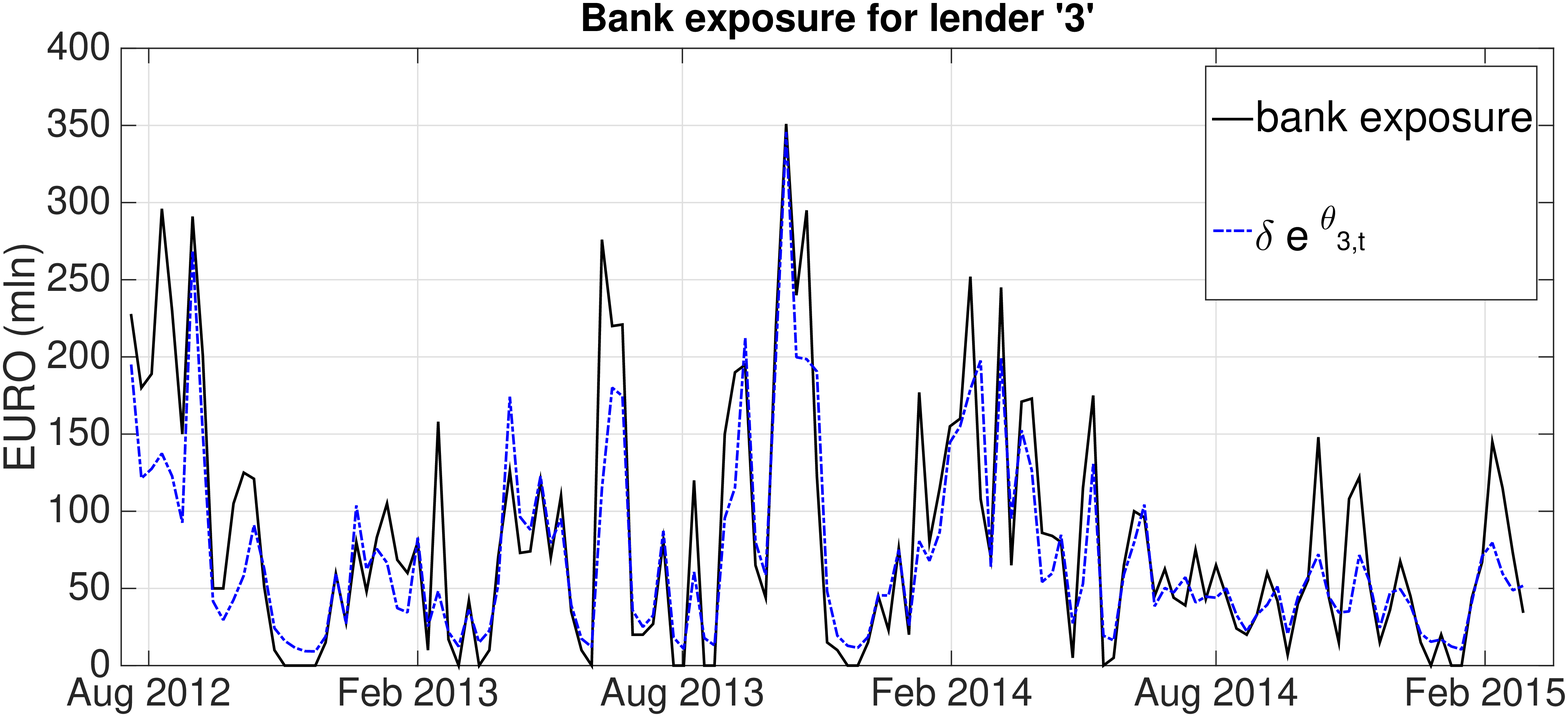}}
\caption{Density estimation of cross sectional (top left panel) and temporal (top right panel) Spearman correlation between the inferred $x_i^{t,out(in)}\equiv e^{\theta^{t,out(in)}_{i}}$ and the corresponded bank exposure $s_i^t$ in e-MID. Bottom panel: an example of time-varying fitness compared with the bank exposure for node `3'. The parameter $\delta$ is chosen in such a way that the maxima of the two time series correspond.}
\label{fit_vs_exposure}
\end{figure}

Node fitness is a latent variable whose time evolution is not observed but inferred according to models of temporal networks. However it is interesting to ask if there exists an observable quantity correlated with it. We show that for the considered dataset, node fitness is correlated with the bank exposure in the e-MID interbank market\footnote{Exposure of bank $i$ is defined as the strength of node $i$ in the weighted network. We refer to it as $s_i^{t,out(in)}$ for generic node $i$. The node out-strength corresponds to the bank asset exposure in e-MID while the node in-strength to the liability.}. 

In fact we observe that the quantity $x_i^{t,out(in)}\equiv e^{\theta^{t,out(in)}_{i}}\:\:\:\forall i=1,...,N$ estimated on data for both TGRG and DAR-TGRG models is strongly correlated with the corresponding bank's exposure in e-MID for the considered week $t$, see the top left panel in Figure \ref{fit_vs_exposure}. This result suggests that, at a given time, banks with larger exposures are the nodes with larger fitnesses $\theta_i^{t,out(in)}$ or equivalently with larger degrees. Furthermore, the time-varying fitness of a node is correlated significantly with its bank exposure (see the top right panel of Figure \ref{fit_vs_exposure}). Finally, in the bottom panel of Figure \ref{fit_vs_exposure} we show an example of this behavior for node 3 whose correlation coefficient is $\rho_{x_3^t,s_3^t}\approx 0.90$. Thus the dynamic fitness model can be seen as a procedure allowing to have some insights on bank exposures having only information on the binary network.

\subsection{Link stability and preferential trading in e-MID}
For credit networks like e-MID, the preferential linkage mechanism reflects the presence of banks which trade preferentially each others. Preferential trading between banks can be detected by comparing empirically observed trading relationships with a null hypothesis that assumes random trading. Hatzopoulos et al. \cite{tumminiello2015} have introduced a statistical test to assess the statistical significance of the observed interbank transactions in order to reveal preferential credit relationships among banks. We apply the same statistical test to show that preferential trading relations in e-MID are associated with link stability.

Following \cite{tumminiello2015} we apply the test to the weekly aggregated e-MID data split in time windows of 3-maintenance periods\footnote{The period of time in which credit institutions have to comply with the minimum reserve requirements is called the reserve maintenance period.  Each reserve maintenance period is equivalent to one calendar month and we aggregate the maintenance periods in groups of three. Hence, we consider 12 3-maintenance periods ranging from April $2^{nd}$, 2012 to February $27^{th}$, 2015.} In each time window and for each link $(i,j)$  we count the number of times $n^{lb}_{ij}$ bank $j$ borrowed money from bank $i$. Then, let $n^l_i$ be the number of times bank $i$ lent money to any other banks and let $n^b_j$ be the number of times bank $j$ borrowed money from any other bank. Finally, let us define $N_T$ as the total number of trades among banks in the system for the considered 3-maintenance period. Under the null hypothesis of random trading, $n^{lb}_{ij}$ follows the hypergeometric distribution $H\left(n^{lb}_{ij}|n^l_i,n^b_j,N_T\right)$.
Hatzopoulos et al. associate preferential trading with  over-expressed number of links with respect to the null hypothesis of random trading, i.e. they use the hypergeometric distribution to associate a p-value with the observed number $n^{lb}_{ij}$. Preferential trading relations $i\rightarrow j$ are the ones rejected according to the statistical test, i.e. with a p-value smaller than the threshold value $\frac{0.05}{a}$ where $a$ is the Bonferroni correction to avoid a large number of false positive validated links because of the multiple hypothesis testing (see \cite{tumminiello2015} for more details).

\begin{figure}[t]
\centering
{\includegraphics[width=0.49\textwidth]{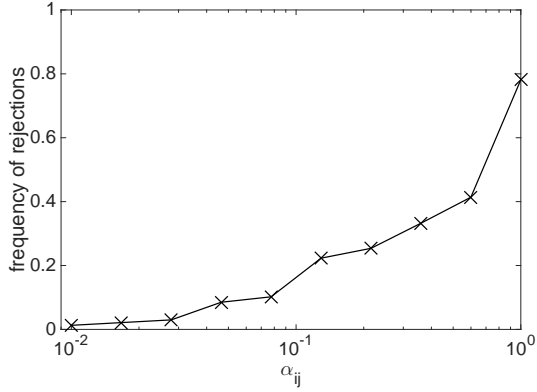}}
\caption{Fraction of statistically validated links according to the test in Hatzopoulos et al. \cite{tumminiello2015} conditional to the value of the estimated $\alpha_{ij}$ parameter measuring the link persistence in the TGRG-DAR model.}
\label{alphaAnalysis}
\end{figure}

Figure \ref{alphaAnalysis} shows the frequency of rejection for the statistical test of Ref. \cite{tumminiello2015} conditional to the estimated $\alpha_{ij}$ parameter measuring the link persistence in the TGRG-DAR model. The clear monotonic behavior indicates that link stability is statistically associated with preferential trading detected according to \cite{tumminiello2015}.

\subsection{Forecasting links}
Finally we compare the proposed network models in their out-of-sample link forecasting performance. We use the first $106$ weekly network observations for model estimation and the last $50$ as our out-of-sample period. In the training phase we estimate the static parameters for the three models and then we adopt the following forecast scheme based on on-line inference. Rolling over the out-of-sample period, at each week $t$ we use the new observed snapshot $\A^t$ to infer the expected $\hat{\Theta}^t$ via Eq. \ref{argmaxtheta}. Then, for DAR-TGRG model we produce the one-step-ahead forecast for each link as
\begin{equation}\label{forecastDARTGRG}
\begin{split}
&\E[A_{ij}^{t+1}|A_{ij}^t,\hat{\theta}_i^t,\hat{\theta}_j^t]=\int d\theta_i^{t+1}d\theta_j^{t+1}\P[A_{ij}^{t+1}=1|A_{ij}^t,\theta_{i}^{t+1},\theta_j^{t+1}]n(\theta_i^{t+1}|\hat{\theta}_i^{t})n(\theta_j^{t+1}|\hat{\theta}_j^{t})=\\
&=\alpha_{ij}A_{ij}^t+(1-\alpha_{ij})\int_0^{\infty}\frac{d\omega}{2} p_{PG}(\omega)\frac{e^{\frac{-4\omega(\phi_{0,i}+\phi_{1,i}\hat{\theta}_{i}^{t}+\phi_{0,j}+\phi_{1,j}\hat{\theta}_{j}^{t})^2+(\sigma_i^2+\sigma_j^2)+4(\phi_{0,i}+\phi_{1,i}\hat{\theta}_{i}^{t}+\phi_{0,j}+\phi_{1,j}\hat{\theta}_{j}^{t})}{8(1+\omega(\sigma_i^2+\sigma_j^2))}}}{\sqrt{1+\omega(\sigma_i^2+\sigma_j^2)}},
\end{split}
\end{equation}
where we have applied the result of Polson et al. \cite{polson2013} as before. The one-step-ahead forecast for the TGRG model is simply obtained by putting $\alpha_{ij}$ equal to $0$ in Eq. \ref{forecastDARTGRG}. The one-step-ahead forecast for DAR(1) model is a standard result of time series analysis given by
\begin{equation}\label{forecastDAR}
\E[A_{ij}^{t+1}|A_{ij}^t]=\alpha_{ij}A_{ij}^t+(1-\alpha_{ij})\chi_{ij}.
\end{equation}
Finally, we compare the Receiving Operating Characteristic (ROC) curves obtained for the three network models (see \cite{friedmanLearningBook} for the definition of ROC curve).
\begin{figure}[t]
\centering
{\includegraphics[width=0.49\textwidth]{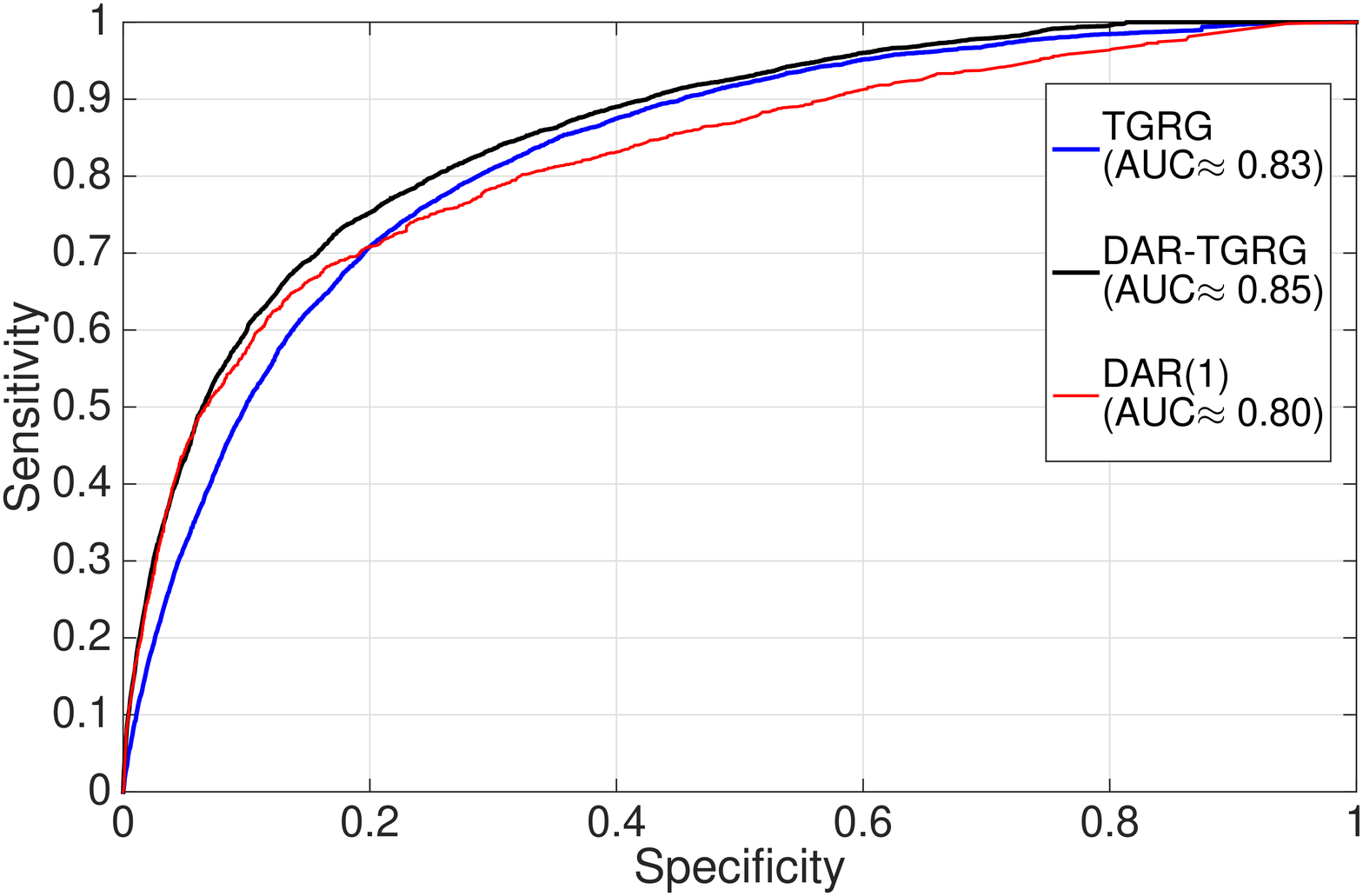}}
{\includegraphics[width=0.49\textwidth]{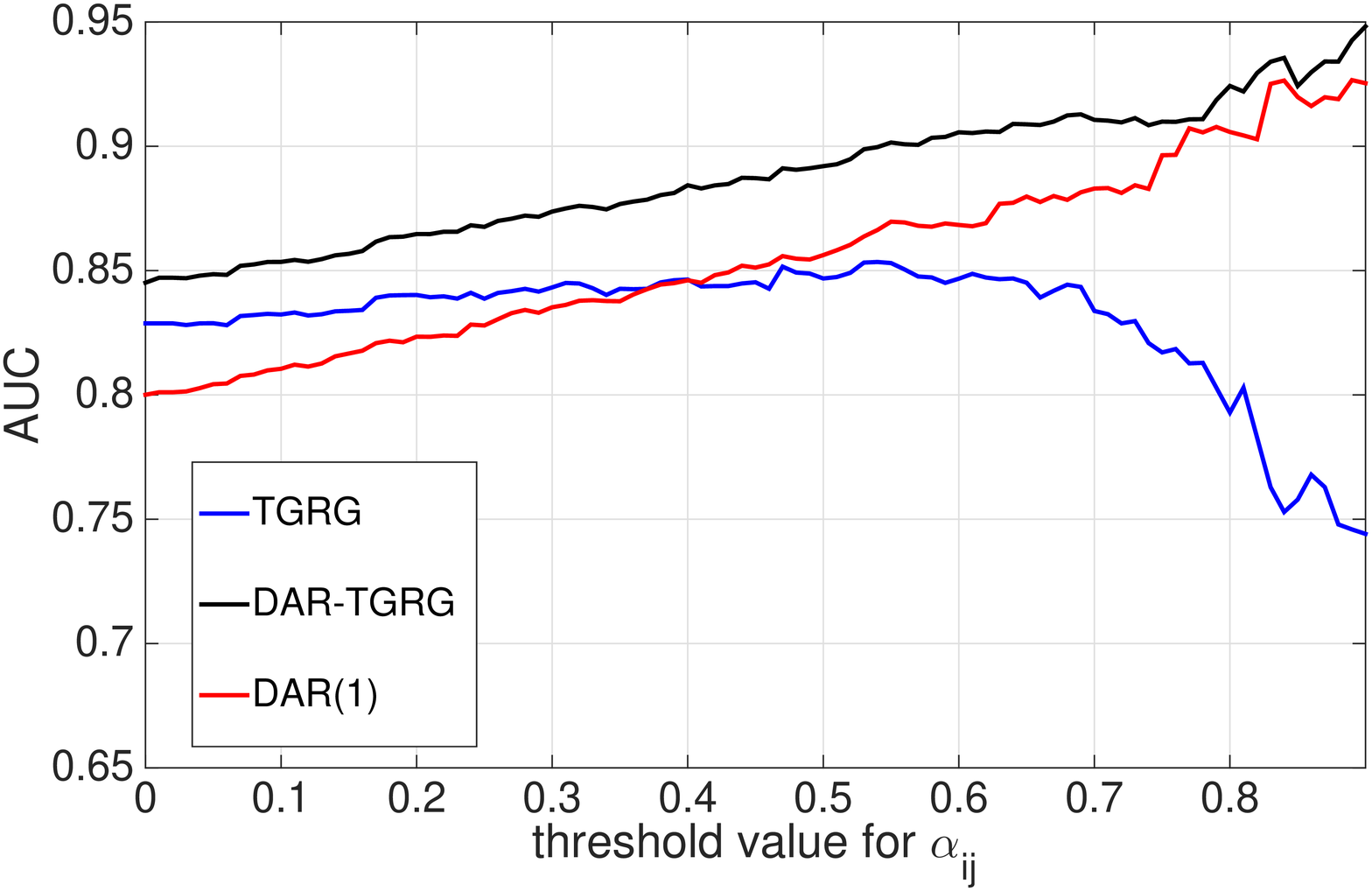}}
\caption{Left panel: ROC curve drawn according to the out-of-sample forecasting exercise: TGRG (blue line), DAR-TGRG (black line) and DAR(1) (red line). Right panel: Area Under the Curve (AUC) of the three models as a function of the threshold value for $\alpha_{ij}$ inferred according to DAR-TGRG.}
\label{roc}
\end{figure}

The results are summarized in Figure \ref{roc}. In the left plot, we compare the three ROC curves and we can notice how the DAR-TGRG model (slightly) outperforms the other models. Furthermore, in the right plot we show the area under the curve (AUC) as a function of a threshold for $\hat{\alpha}_{ij}$ estimated according to DAR-TGRG model. In other words, we compare the AUC considering only the links for which the $\hat{\alpha}_{ij}$ estimated by the the DAR-TGRG model is larger than a threshold value. We find that taking into account both of fitness dynamics and preferential linkage better forecast links, i.e. DAR-TGRG outperforms always the other models. When we consider links with both high and low persistence, the TGRG model outperforms the DAR(1) network model, that is the evolution of the network topology is more important than preferential linkage in determining the average characteristics of the e-MID network. However, the link copying mechanism associated with the DAR(1) model characterizes better than the fitness dynamics the persistence pattern associated with a smaller set of links representing the preferential relations among banks. In fact, there exists a value of the threshold (around $0.4$) after which the AUC associated with the DAR(1) model is larger than the one for TGRG.

\section{Conclusions}\label{conclusion}

In this paper we introduce a novel state-of-the-art statistical methodology to describe link persistence and fitness dynamics in temporal networks. We model a Markov dynamics for both observed and unobserved time-varying states which drive the evolution of the network. The analytic tractability of the autoregressive network ensemble we propose allows us to easily calibrate our parameters from the data with a general likelihood maximization iterative procedure. The introduction of the autoregressive dynamics permits link forecasting by taking account of memory properties of the network system. Then, the estimation method we introduce allows online-inference of the time-varying parameters which is particularly useful from a computational point of view to face the problem of link prediction. 

The contribution of the paper is twofold. First, the introduction of autoregressive endogenous components displays the clear advantage of describing the network evolution via time-varying states which reproduce the network topology as well as capturing the local property of link persistence, thus going beyond a single snapshot analysis where parameters are chosen for each network snapshot, independently. Second, the analysis on real data from the eMID interbank network from 2012 to 2015 (weekly aggregated) displays the statistical equivalence between link stability, identified by positive value of the persistence parameter, and preferential trading, identified by over-expressed number of trades between counterparties. Hence, our methodology permits to disentangle preferential trading from random trading in dynamic trading networks such as the eMID money market. Finally, the forecasting performance of the model points out both fitness dynamics and link persistence as linkage mechanisms in the process of network formation for the credit market.

As future outlooks, the formalism discussed in the paper could be also be applied to more general memory kernel function of the autoregressive model governing the evolution of the system as well as it could permit the introduction of exogenous factors driving the fitness dynamics or the local link probability. Furthermore, a challenging issue is the introduction of a dependence structure for the dynamic fitnesses. We also note that the estimation method we introduce to obtain our results for dynamic networks is quite general, and could be used to obtain similar results for other types of fitness dynamics.

\section*{Acknowledgment}
We acknowledge financial support from the grant SNS16LILLB ``Financial networks: statistical models, inference, and shock propagation''. FL acknowledges support by the European Community's H2020 Program under the scheme INFRAIA-1- 2014-2015: Research Infrastructures, grant agreement no. 654024 SoBigData: Social Mining \& Big Data Ecosystem.

\appendix

\section{ Two-point distribution function for TGRG} 
In the TGRG model, the node fitness is autocorrelated in time when $\phi_{i,1}\neq 0$. An autocorrelated fitness reflects the autocorrelation of the degree and ultimately of all links incident to the node\footnote{This is always true in the case of finite network size.}. A positive autocorrelated fitness is associated with persistence of links. This effect can be characterized by studying the two-point distribution function or equivalently the autocorrelation function. In the TGRG model we can compute semi-analytically the the two-point distribution function,
\begin{equation}\label{twopointf}
\begin{split}
&\P(A_{ij}^t=1,A_{ij}^{t-\tau}=1)=\int d\theta_i^td\theta_j^td\theta_i^{t-\tau}d\theta_j^{t-\tau}\P(A_{ij}^t=1|\theta_i^t,\theta_j^t)\P(A_{ij}^{t-\tau}=1|\theta_i^{t-\tau},\theta_j^{t-\tau})p(\theta_i^t,\theta_i^{t-\tau})p(\theta_j^t,\theta_j^{t-\tau})=\\
&=\int d\theta_i^{t-\tau}d\theta_j^{t-\tau}\frac{1}{1+e^{-(\theta_i^{t-\tau}+\theta_j^{t-\tau})}}n(\theta_i^{t-\tau})n(\theta_j^{t-\tau})\times\\
&\times\int \left[\prod_{q=1}^{\tau-1}\prod_{a=i,j}n(\theta_a^{t-\tau+q}|\theta_a^{t-\tau+(q-1)})d\theta_a^{t-\tau+q}\right] \int d\theta_i^t d\theta_j^t\frac{1}{1+e^{-(\theta_i^t+\theta_j^t)}}n(\theta_i^t|\theta_i^{t-1})n(\theta_j^t|\theta_j^{t-1})=\\
&=\int d\theta_i^{t-\tau}d\theta_j^{t-\tau}\frac{1}{1+e^{-(\theta_i^{t-\tau}+\theta_j^{t-\tau})}}n(\theta_i^{t-\tau})n(\theta_j^{t-\tau})\int_0^{\infty}\frac{d\omega}{2} p_{PG}(\omega)K^{\tau}(\omega|\theta_i^{t-\tau},\theta_j^{t-\tau})
\end{split}
\end{equation}
where we have applied the result of Polson et al. \cite{polson2013} as before and
$$
K^\tau(\omega|\theta_i^{t-\tau},\theta_j^{t-\tau})=\frac{e^{\frac{-4\omega(\mu^\tau_i+\mu^\tau_j)^2+((\sigma^\tau_i)^2+(\sigma^\tau_j)^2)+4(\mu^\tau_i+\mu^\tau_j)}{8(1+\omega((\sigma^\tau_i)^2+(\sigma^\tau_j)^2))}}}{\sqrt{1+\omega((\sigma^\tau_i)^2+(\sigma^\tau_j)^2)}}
$$
with
$$
\mu_a^\tau=\phi_{0,a}\left(\sum_{t=0}^{\tau-1}(\phi_{1,a})^t\right)+(\phi_{1,a})^\tau\theta_a^{t-\tau}\:\:\:a=i,j
$$
$$
(\sigma_a^\tau)^2=\sigma_a^2(\sum_{t=0}^{\tau-1}(\phi_{1,a}^2)^t)\:\:\:a=i,j.
$$
The last recursive formulas are obtained by integrating over the Gaussian transition probabilities in Eq. \ref{twopointf}. Let us notice that $\mu_a^\tau$ and $(\sigma_a^\tau)^2$ converge to the mean and the variance of the marginal distribution for $\theta_a^t$ in the limit $\tau\rightarrow\infty$ as we can expect for the standard AR(1) process.

Then, the two-point distribution function can be obtained by integrating over the Gaussian marginals, i.e. $n(\theta_i^{t-\tau})$ and $n(\theta_j^{t-\tau})$, and finally by performing the numerical integration over the probability density function associated with the Polya-Gamma distribution. Let $\tilde{\mu}_a\equiv\frac{\phi_{0,a}}{1-\phi_{1,a}}$ and $\tilde{\sigma}^2_a\equiv\frac{\sigma_a^2}{1-\phi_{1,a}^2}$ $a=i,j$ be the mean and the variance of the Gaussian marginal distribution for $\theta_a^{t-\tau}$. It is
\begin{equation}\label{tf_EAtau}
{\footnotesize
\begin{split}
&\P(A_{ij}^t=1,A_{ij}^{t-\tau}=1)=\int_0^\infty\frac{d\omega}{2}p_{PG}(\omega)\int_0^\infty\frac{d\zeta}{2}p_{PG}(\zeta)\times\\
&\times \frac{e^{f(\omega,\zeta,\phi_{0,i},\phi_{0,j},\phi_{1,i},\phi_{1,j},\sigma_i,\sigma_j)}}{\sqrt{1+\zeta(\tilde{\sigma}_i^2+\tilde{\sigma}_j^2)+\omega(C_i^\tau\sigma_i^2+C_j^\tau\sigma_j^2+(B_i^\tau)^2\tilde{\sigma}_i^2+(B_j^\tau)^2\tilde{\sigma}_j^2)+\zeta\omega(\tilde{\sigma}_i^2(C_i^\tau\sigma_i^2+C_j^\tau\sigma_j^2)+\tilde{\sigma}_j^2(C_i^\tau\sigma_i^2+C_j^\tau\sigma_j^2)+\tilde{\sigma}_i^2\tilde{\sigma}_j^2(B_i^\tau-B_j^\tau)^2)}}
\end{split}
}
\end{equation}
where
{\scriptsize
$$
f(\omega,\zeta,\bm{\Phi}_i,\bm{\Phi}_j)=\frac{N(\omega,\zeta,\bm{\Phi}_i,\bm{\Phi}_j)}{8\left(1+\zeta(\tilde{\sigma}_i^2+\tilde{\sigma}_j^2)+\omega(C_i^\tau\sigma_i^2+C_j^\tau\sigma_j^2+(B_i^\tau)^2\tilde{\sigma}_i^2+(B_j^\tau)^2\tilde{\sigma}_j^2)+\zeta\omega(\tilde{\sigma}_i^2(C_i^\tau\sigma_i^2+C_j^\tau\sigma_j^2)+\tilde{\sigma}_j^2(C_i^\tau\sigma_i^2+C_j^\tau\sigma_j^2)+\tilde{\sigma}_i^2\tilde{\sigma}_j^2(B_i^\tau-B_j^\tau)^2)\right)}
$$}
and
$$
{\scriptsize
\begin{split}
N(\omega,\zeta,\bm{\Phi}_i,\bm{\Phi}_j)=&4(A^{\tau}_i+A^{\tau}_j)+4(1+B^\tau_i)\tilde{\mu}_i+4(1+B^\tau_j)\tilde{\mu}_j+C^\tau_i\sigma_i^2+C^\tau_j\sigma_j^2+(1+B^\tau_i)^2\tilde{\sigma}_i^2+(1+B^\tau_j)^2\tilde{\sigma}_j^2+\\
+\zeta&\:\:(\tilde{\sigma}_i^2\tilde{\sigma}_j^2(B^\tau_i-B^\tau_j)^2 -4(\tilde{\mu}_i+\tilde{\mu}_j)^2+\\
&+\tilde{\sigma}_i^2\:(4A^\tau_i+4A^\tau_j+4\tilde{\mu}_j(B^\tau_j-B^\tau_i)+C^\tau_i\sigma_i^2+C^\tau_j\sigma_j^2)+\\
&+\tilde{\sigma}_j^2(4A^\tau_i+4A^\tau_j+4\tilde{\mu}_i(B^\tau_i-B^\tau_j)+C^\tau_i\sigma_i^2+C^\tau_j\sigma_j^2))+\\
-\omega&\:\:(4(A^\tau_i+A^\tau_j)^2+4(B^\tau_i\tilde{\mu}_i+B^\tau_j\tilde{\mu}_j)(2A^\tau_i+2A^\tau_j+B^\tau_i\tilde{\mu}_i+B^\tau_j\tilde{\mu}_j)+4(\tilde{\mu}_i+\tilde{\mu}_j)(-1+\zeta(\tilde{\mu}_i+\tilde{\mu}_j))(C^\tau_i\sigma_i^2+C^\tau_j\sigma_j^2)+\\
&+\tilde{\sigma}_i^2(4(A^\tau_i+A^\tau_j+(B^\tau_j-B^\tau_i)\tilde{\mu}_j)(B^\tau_i(1-\zeta\tilde{\mu}_j)+\zeta(A^\tau_i+A^\tau_j+B^\tau_j\tilde{\mu}_j))-(C^\tau_i\sigma_i^2+C^\tau_j\sigma_j^2))+\\
&+\tilde{\sigma}_j^2(4(A^\tau_i+A^\tau_j+(B^\tau_i-B^\tau_j)\tilde{\mu}_i)(B^\tau_j(1-\zeta\tilde{\mu}_i)+\zeta(A^\tau_i+A^\tau_j+B^\tau_i\tilde{\mu}_i))-(C^\tau_i\sigma_i^2+C^\tau_j\sigma_j^2))+\\
&-\tilde{\sigma}_i^2\tilde{\sigma}_j^2(B^\tau_i-B^\tau_j)^2)
\end{split}}
$$
where we have defined for notational simplicity 
$$
A^{\tau}_a\equiv\phi_{0,a}\left(\sum_{t=0}^{\tau-1}(\phi_{1,a})^t\right),\:\:\:B^{\tau}_a\equiv(\phi_{1,a})^\tau,\:\:\:C^{\tau}_a\equiv (\sum_{t=0}^{\tau-1}(\phi_{1,a}^2)^t)\:\:\:a=i,j.
$$
Finally, the ACF can be obtained by noticing that $\E[A_{ij}^{t}A_{ij}^{t-\tau}]\equiv\P(A_{ij}^t=1,A_{ij}^{t-\tau}=1)$ and the unconditional expectation for $A_{ij}^t$ is
$$
\E[A_{ij}^{t}]=\int d\theta_{i,t}d\theta_{j,t}\P(A_{ij}^{t}=1|\theta_{i,t},\theta_{j,t})n(\theta_{i,t})n(\theta_{j,t})=\int \frac{d\omega}{2} p_{PG}(\omega)\frac{e^{\frac{-4\omega(\tilde{\mu}_i+\tilde{\mu}_j)^2+(\tilde{\sigma}_i^2+\tilde{\sigma}_j^2)+4(\tilde{\mu}_i+\tilde{\mu}_j)}{8(1+\omega(\tilde{\sigma}_i^2+\tilde{\sigma}_j^2))}}}{\sqrt{1+\omega(\tilde{\sigma}_i^2+\tilde{\sigma}_j^2)}}.
$$

\end{document}